\documentclass[lettersize,journal]{IEEEtran}
\usepackage{algorithmic}
\usepackage{array}
\usepackage[caption=false,font=footnotesize]{subfig}
\usepackage{textcomp}
\usepackage{stfloats}
\usepackage{url}
\usepackage{verbatim}
\usepackage{graphicx}
\def\BibTeX{{\rm B\kern-.05em{\sc i\kern-.025em b}\kern-.08em
    T\kern-.1667em\lower.7ex\hbox{E}\kern-.125emX}}
\usepackage{cite}
\usepackage{amsmath,amssymb,amsfonts}
\usepackage{xcolor}
\usepackage{algorithm,algorithmic}
\usepackage{hyperref}
\hypersetup{hidelinks=true}
\usepackage{textcomp}
\usepackage{booktabs}
\usepackage{tabularx}
\usepackage{makecell}
\usepackage{multirow}
\usepackage{array}   
\begin{document}
\bstctlcite{IEEEexample:BSTcontrol}
\title{US-Seg: One Reference for Universal Ultrasound Segmentation}
\author{Chen Ma, \IEEEmembership{Graduate Student Member, IEEE}, Yunshu Li, Shuyu Liang, Qin Wang, Junhu Fu, Zeju Li, Yi Guo, \IEEEmembership{Member, IEEE}, and Yuanyuan Wang, \IEEEmembership{Senior Member, IEEE}
\thanks{This work was supported by National Key R\&D Program of China (2024YFF0507300, 2024YFF0507303) and the National Natural Science Foundation of China (62531004). (Corresponding authors: Yi Guo; Yuanyuan Wang)}
\thanks{Chen Ma, Yunshu Li, Shuyu Liang, Qin Wang, Junhu Fu, Zeju Li, Yi Guo, and Yuanyuan Wang are with the College of Biomedical Engineering, Fudan University, Shanghai 200433, China (email: \{cma24, ysli24, syliang22, wangqin23, jhfu21\}@m.fudan.edu.cn;
\{zejuli, guoyi, yywang\}@fudan.edu.cn).}}

% The paper headers
\markboth{Journal of \LaTeX\ Class Files,~Vol.~14, No.~8, August~2021}%
{Shell \MakeLowercase{\textit{et al.}}: A Sample Article Using IEEEtran.cls for IEEE Journals}

\maketitle

\begin{abstract}
Recent ultrasound foundation models have demonstrated strong feature extraction capabilities across organs, tasks, and imaging conditions. However, converting these representations into segmentation still requires finetuning on annotated target data. To reduce reliance on task-specific finetuning, existing universal ultrasound segmentation methods typically adopt prompt-based target specification, mainly through spatial prompts in SAM-style models or task prompts exemplified by MOFO. Yet, directly segmenting targets unseen during training still requires repeated per-image interaction or target-specific adaptation. We propose US-Seg, a universal ultrasound segmentation framework that uses only a single annotated reference to specify the target and directly predicts its mask in query images without finetuning. The fundamental insight of US-Seg is to reformulate segmentation from direct image-to-label mapping for predefined targets into a shared structural recovery process learned from heterogeneous reference-query pairs through masked label reconstruction, thereby making a single reference sufficient for target specification. Boundary-centric label masking emphasizes ambiguous boundaries, encouraging transferable structural cues rather than target-specific appearance patterns. Using these structure-enriched representations, multi-level bidirectional reference-query interaction aligns the reference-defined target with the query. Finally, a prediction-aware iterative decoder uses intermediate predictions to progressively correct residual structural errors under reference guidance. Without repeated interaction or laborious annotation, a single US-Seg model with only one reference achieves an average Dice of 81.85\% across six evaluations covering cross-site and cross-task generalization, outperforming the strongest label-reference method by 24.30\%, while also delivering the best in-domain performance. These results advance universal ultrasound segmentation toward direct, scalable application across evolving clinical targets and diverse imaging environments. 
\end{abstract}


\section{Introduction}
\label{sec:introduction}
\IEEEPARstart{R}{ecent} ultrasound foundation models, exemplified by USFM~\cite{usfm}, have demonstrated that large-scale pretraining on heterogeneous ultrasound data can learn transferable representations across diverse organs, imaging conditions, and downstream tasks. These advances have established a promising foundation for universal ultrasound image analysis by reducing the dependence on task-specific feature learning. However, representation-level universality does not directly translate into segmentation-level universality, as existing ultrasound segmentation systems still typically require task-specific decoders, annotated target-domain data, or additional finetuning to instantiate the learned representations for each new anatomical target or imaging domain~\cite{tip0}. Consequently, a critical gap remains between general-purpose ultrasound representation learning and direct universal segmentation, particularly when the imaging domain or anatomical target is unseen during training.

To move toward segmentation-level universality, recent universal segmentation methods have introduced flexible mechanisms for target specification. Spatial-prompted methods specify the target through location cues such as points and boxes, as exemplified by SAM~\cite{sam} and its medical variant MedSAM~\cite{medsam}, while SAMUS~\cite{samus} and UltraSAM~\cite{ultrasam} further adapt this paradigm to ultrasound, achieving strong segmentation performance across diverse targets. Such spatial prompts mainly convey target location, provide limited structural and semantic information, and typically require repeated prompting for each query image. In contrast, task-prompted methods, such as MOFO~\cite{mofo} and PerceptGuide~\cite{perceptguide}, specify the segmentation target at the task level and support multiple organs or tasks within a unified model. Their target space remains defined during training, limiting direct extension to previously unseen anatomical structures. A more flexible alternative is to define the target through annotated references, as explored by UniverSeg~\cite{universeg} and INSID3~\cite{insid3}. Compared with spatial and task prompts, annotated references provide richer structural and semantic cues, support target specification beyond fixed categories, and can be reused across multiple queries. If the target can be reliably defined from a single annotation, a reusable reference could provide flexible and consistent specification across queries without redefining a fixed task space or repeating per-image specification.

\begin{figure*}[htbp]
    \centering
    \includegraphics[width=0.95\linewidth]{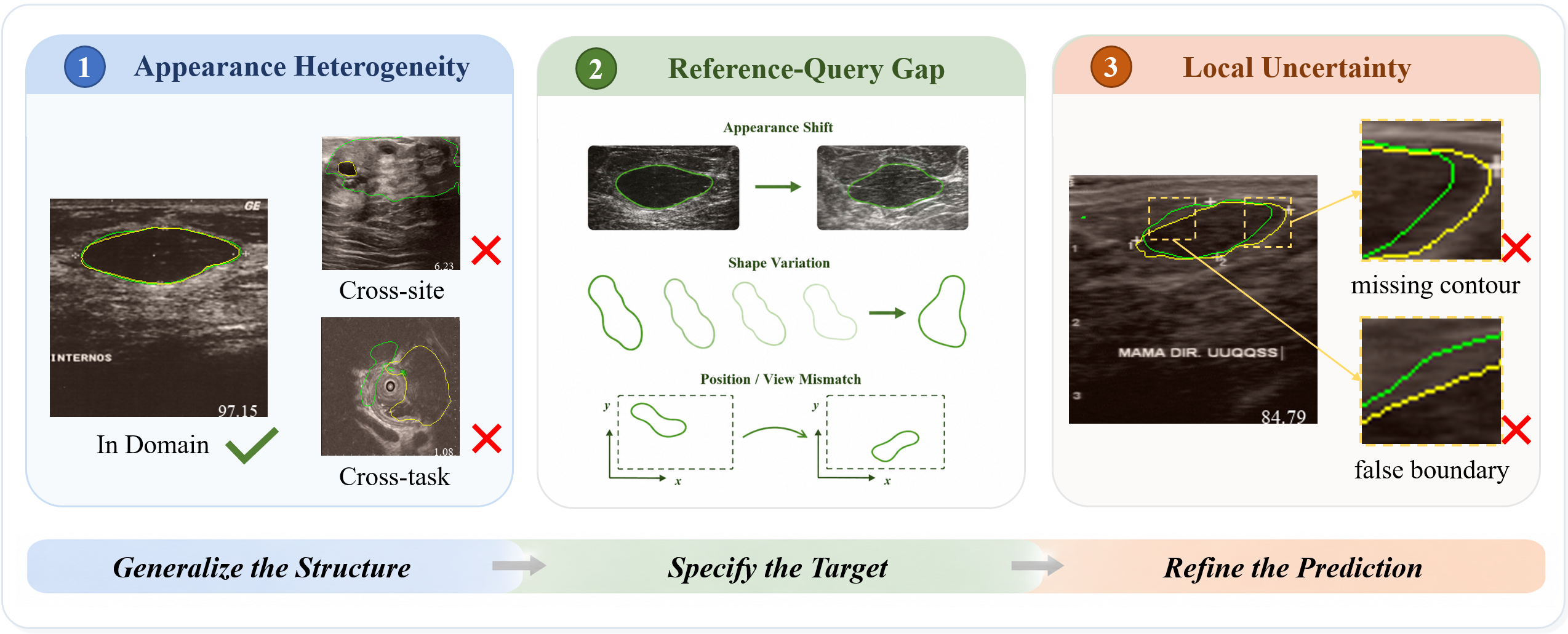}
    \caption{Key challenges in single-reference universal ultrasound segmentation.}
    \label{fig1}
\end{figure*}

Despite this potential, making a single annotated reference sufficient for universal ultrasound segmentation remains particularly challenging, posing three progressively connected challenges. First, how to \textbf{generalize the structure} across heterogeneous ultrasound distributions. Compared with more standardized medical imaging modalities, ultrasound is particularly susceptible to appearance variations across devices, operators, and acquisition angles. As a result, the same anatomical structure can appear markedly different across imaging conditions, making appearance-dependent correspondence unreliable and hindering the generalization of stable structural knowledge across domains~\cite{tip1,tip2}. Second, given such transferable structural knowledge, how to \textbf{specify the target} from only one annotated reference despite diverse variations across the reference-query gap. Existing reference-based methods often alleviate large appearance discrepancies by using multiple annotated references or support examples, sometimes requiring dozens of reference cases to provide sufficient coverage~\cite{universeg, tyche}. However, relying on many annotated references reintroduces substantial annotation burden and undermines the practical motivation of reference-based segmentation. A universal ultrasound model should instead minimize annotation burden, making a single reusable reference sufficient for reliable target specification across queries. Third, once the target is specified, how to \textbf{refine the prediction} under local structural ambiguity. Existing methods may approximately localize the intended structure, but weak boundaries, missing contours, false edges, acoustic shadows, and low-contrast regions often prevent accurate segmentation of its detailed shape~\cite{wang2022boundary, tip3}. Importantly, the annotated reference provides not only an initial cue for target identification but also a reusable structural guide for correcting uncertain regions. Fully exploiting this information requires progressive reference-guided refinement rather than one-shot prediction.

To address these challenges, we propose US-Seg, a universal ultrasound segmentation framework that reframes conventional image-to-label prediction as reference-guided reconstruction over heterogeneous reference-query pairs, enabling reliable segmentation with only a single reusable reference. During training, US-Seg learns to recover masked label structures from heterogeneous reference-query pairs, encouraging shared structural reconstruction across different anatomical targets and imaging distributions. At inference, given only a single annotated reference image-mask pair and a query image, US-Seg can directly segment all query images in a dataset without any target-specific adaptation. The annotated reference defines what to segment by providing the target identity together with its structural and appearance context. This formulation enables a single model to transfer segmentation across imaging sites and instantiate new segmentation tasks for an entire dataset from just one annotated reference. The main contributions of this work are summarized as follows:

\begin{itemize}
\item[$\bullet$] We introduce US-Seg, a reference-guided reconstruction framework for universal ultrasound segmentation that enables a single model to flexibly segment both observed and previously unseen anatomical targets specified by only one reusable annotated reference.  
\item[$\bullet$] Using heterogeneous reference-query pairs as the learning unit, we develop a unified reconstruction design that generalizes the structure by learning target-agnostic structural recovery through boundary-centric label masking, specifies the target by establishing multi-level bidirectional reference-query interaction with the annotated reference, and refines the boundary by progressively correcting residual structural errors through prediction feedback and reference-conditioned guidance, enabling reliable segmentation from a single reusable reference.
\item[$\bullet$] We conduct extensive experiments across diverse ultrasound benchmarks, demonstrating strong finetuning-free cross-site and cross-task generalization of US-Seg, while preserving state-of-the-art performance on in-domain distributions.
\end{itemize}

\section{Related Work}
\subsection{Universal Representation Learning}
Universal representation learning has emerged as an important paradigm in medical image analysis, aiming to learn transferable features from large-scale heterogeneous data that can support a wide range of clinical tasks~\cite{A1,A2}. Generalist medical AI and medical foundation models have further demonstrated the value of reusable representations across organs, modalities, and tasks~\cite{A3,A4,A5}. This paradigm is particularly relevant to ultrasound, where image appearance varies substantially across anatomical structures, probe positions, operators, imaging protocols, and device vendors. USFM~\cite{usfm} established a universal ultrasound representation backbone through multi-organ, multi-center, and multi-device pretraining, demonstrating strong transferability across label-efficient classification, segmentation, and enhancement tasks. URFM~\cite{A6} further advanced ultrasound representation learning through representation-based masked image modeling tailored to the low signal-to-noise characteristics of ultrasound images. TinyUSFM~\cite{A7} extended this direction toward efficiency, showing that compact ultrasound foundation models can retain versatility across organs and tasks through carefully designed distillation. Together, these studies establish universal representation learning as an important foundation for scalable ultrasound image analysis.

Despite these advances, existing ultrasound foundation models primarily focus on learning transferable representations and generally serve as reusable encoders. For downstream segmentation, they still require task-specific decoders, supervised adaptation, or target-domain finetuning. More importantly, transferable features alone do not ensure that anatomical structure can be consistently recovered across heterogeneous ultrasound distributions, leaving structural generalization at the segmentation level insufficiently addressed.

\subsection{Promptable Segmentation}
Promptable segmentation advances universal segmentation by introducing explicit mechanisms for specifying what should be segmented. Spatial-prompted methods define the target through location cues in the query image. SAM~\cite{sam} established this paradigm using points, boxes, and masks, while MedSAM~\cite{medsam} extended it to medical image segmentation. Building on the same promptable paradigm, SAMUS~\cite{samus} and UltraSAM~\cite{ultrasam} further adapt SAM-style segmentation to ultrasound and achieve strong performance across diverse ultrasound targets. Although spatial prompts allow flexible segmentation beyond fixed semantic categories, they require the target to be specified separately for each query image through points, boxes, or other spatial cues. Consequently, repeated prompting is needed across images, leading to recurring user interaction as the number of queries increases. 

Task-prompted methods address target specification from a different perspective by explicitly encoding which organ or segmentation task should be performed. Methods such as MOFO~\cite{mofo} and PerceptGuide~\cite{perceptguide} incorporate task-level guidance, anatomical priors, or perceptual information to support multiple organs or tasks within a unified model. Compared with spatial prompts, task prompts provide a more explicit representation of target identity, but they are limited to segmentation targets defined during training, so that they cannot flexibly specify previously unseen anatomical targets. 

Thus, existing promptable segmentation faces a trade-off between flexible location-based specification and explicit but predefined target identity, motivating target-specification mechanisms that can provide both flexibility and consistent target definition across images.

\subsection{Label-Reference Segmentation} Label-reference segmentation provides flexible target specification by defining the intended structure through annotated examples rather than spatial locations or predefined task identities. SegGPT~\cite{C1} formulates segmentation tasks through contextual image-mask examples, while PerSAM~\cite{C2} further demonstrates single-reference personalization toward a specific visual target. However, reliable target transfer still depends on effective reference-query correspondence, which becomes difficult under substantial appearance discrepancies.

In medical imaging, UniverSeg~\cite{universeg} and Tyche~\cite{tyche} improve target specification through multiple annotated support examples, but at the cost of increased annotation burden. More recent methods such as INSID3~\cite{insid3} move toward one-reference segmentation by transferring the intended target through cross-image correspondence. However, the large appearance variations and structural ambiguities commonly encountered in ultrasound can weaken cross-image correspondence, making reliable target transfer and precise boundary recovery difficult.

Taken together, existing label-reference methods establish the feasibility of flexible target specification, but reliable one-reference segmentation remains challenging under the substantial distributional shifts and structural ambiguities encountered in ultrasound. Therefore, universal ultrasound segmentation requires a reference-guided mechanism that can generalize the structure across heterogeneous distributions, preserve the target identity across the reference-query gap, and continuously exploit reference information to refine uncertain predictions.

\section{Methodology}
\subsection{Method Overview}

\begin{figure*}[htbp]
    \centering
    \includegraphics[width=0.9\linewidth]{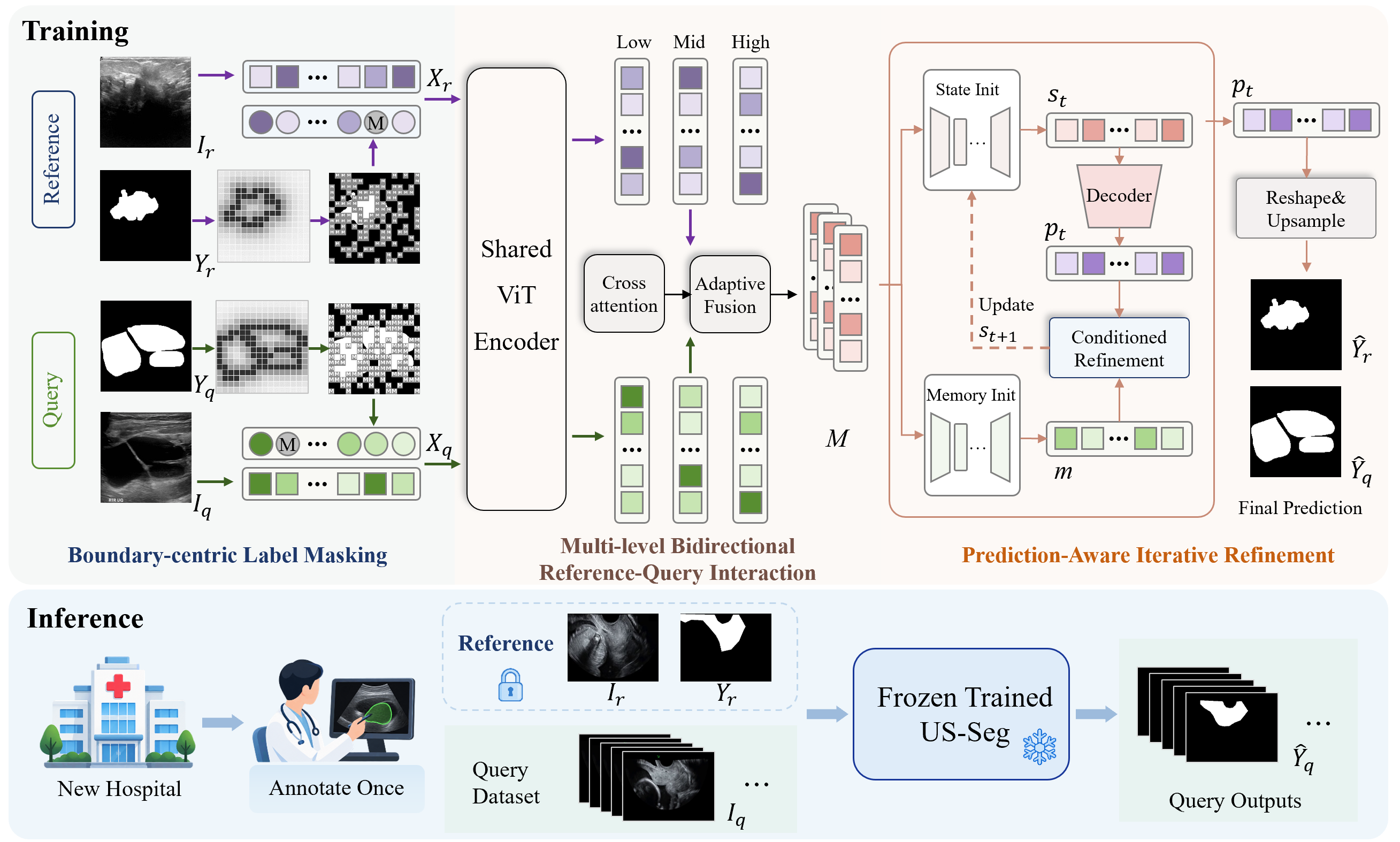}
    \caption{Framework overview of US-Seg.}
    \label{fig2}
\end{figure*}

We propose US-Seg, a universal ultrasound segmentation framework that formulates heterogeneous segmentation tasks as a shared reference-guided reconstruction problem instead of image-label training. Given a reference image $I_{ref}$ with its corresponding annotation $Y_{ref}$, which together define the segmentation target, and a query image $I_{q}$, US-Seg predicts the corresponding query mask $\widehat{Y_{\mathrm{q}}}$. Within each reference-query pair, the reference and query correspond to the same segmentation target but are sampled from different patients. The reference annotation identifies which anatomical structure should be segmented, while the reference image provides its visual, structural, and surrounding anatomical context. During training, the reference and query labels are represented as patch-level label tokens and partially or fully masked for reconstruction, allowing the model to learn shared structural recovery across heterogeneous reference-query pairs. During inference, the query label is predicted directly from the annotated reference pair $(I_{ref}, Y_{ref})$ and the query image $I_q$, without query-side annotation.

US-Seg consists of three components. Boundary-centric label masking preferentially masks structurally ambiguous regions to learn target-agnostic structural recovery across heterogeneous anatomies. Building on this transferable structural representation, multi-level bidirectional reference-query interaction establishes robust correspondence between reference and query through bidirectional cross-branch interaction and adaptive fusion, enabling the target defined by the reference to be identified in the query. Based on this representation, a prediction-aware iterative refinement decoder feeds intermediate predictions back into the decoding process and progressively corrects residual structural errors through gated, reference-conditioned latent-state updates. Together, these components enable direct segmentation of open-ended ultrasound targets without target-specific parameter adaptation.

\subsection{Boundary-centric Label Masking}
\label{sec:blm}

\begin{figure*}[htbp]
    \centering
    \includegraphics[width=0.9\linewidth]{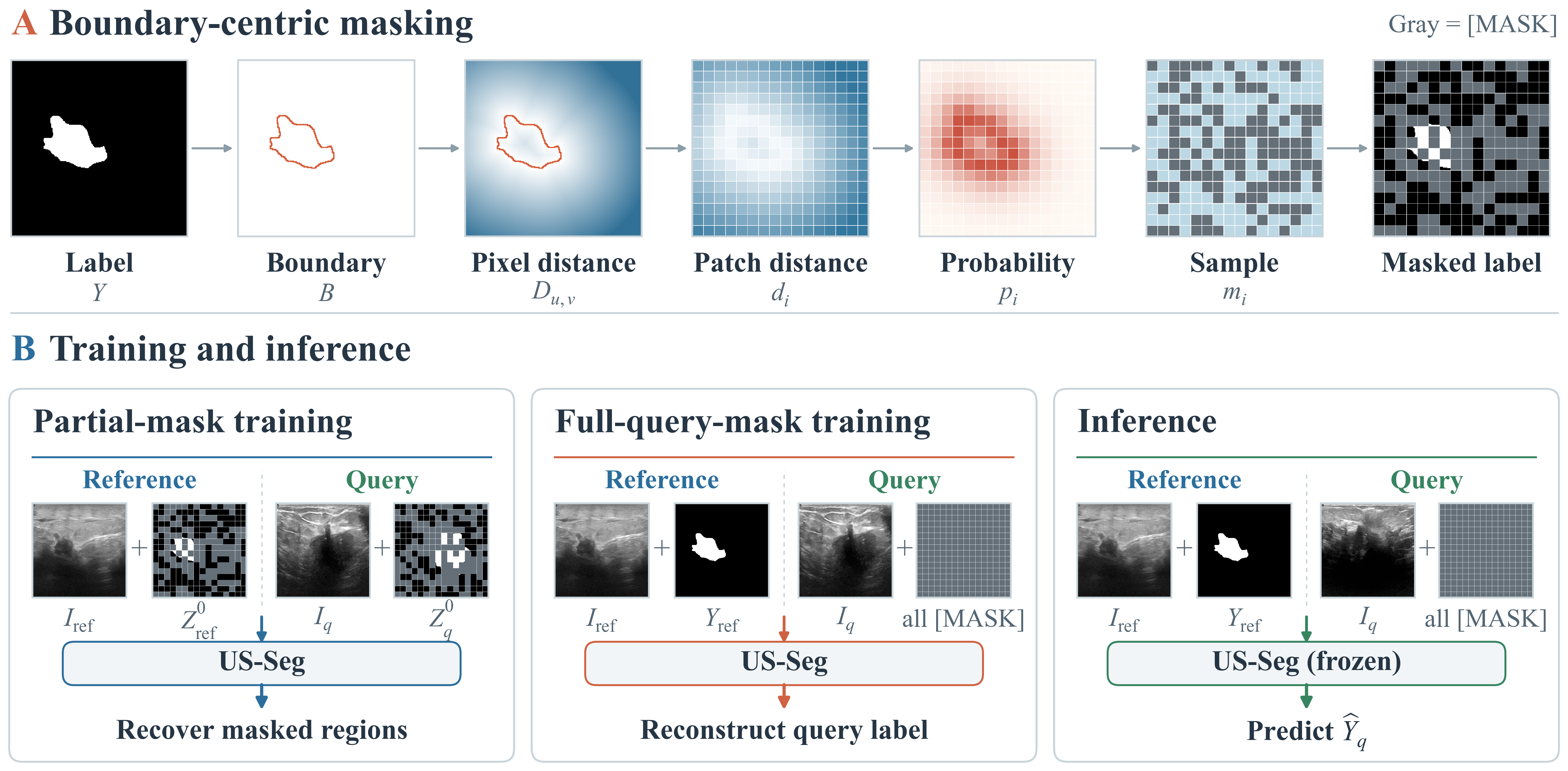}
    \caption{Schematic illustration of boundary-centric label masking.}
    \label{fig2}
\end{figure*}

We introduce boundary-centric label masking to promote structural generalization across heterogeneous ultrasound distributions. Here, we assign higher masking probabilities to label patches near object boundaries, encouraging the model to recover uncertain shape information from stable contextual cues rather than relying on target-specific visual patterns.

To construct the label tokens for both branches, let Y denote the binary annotation mask of either the reference or query image. We first partition Y according to the ViT patch grid~\cite{vit}. For the i-th patch region $P_{i}$, the corresponding label patch is defined as:

\begin{equation}
L_{i}=Vec\left(Y\middle| P_{i}\right),i=1,\ldots ,N. 
\end{equation}

Each visible label patch is mapped into a label token through a learnable mapping $\phi \left(\cdot \right)$, while a masked patch is represented by a learnable mask embedding $e_{m}$:

\begin{equation}
    z_{i}^{\left(0\right)}=\left\{\begin{gathered}e_{m},\ m_{i}=1,\\ \mathrm{\phi }\left(L_{i}\right),m_{i}=0.\end{gathered}\right.
\end{equation}

For each branch $b\in\{ref,q\}$, the image tokens and the corresponding masked label tokens are concatenated along the token dimension to form the joint input sequence $X_{b}$.

To construct the boundary-centric masking distribution, we first extract a pixel-level boundary map from Y using a morphological gradient with a 3×3 structuring element:
\begin{equation}
   B=I\left(Dilate_{3\times 3\! }\left(Y\right)-Erode_{3\times 3\! }\left(Y\right) \right), 
\end{equation}
where $Dilate_{3\times 3}\left(\cdot \right)$ and $Erode_{3\times 3}\left(\cdot \right)$ denote binary dilation and erosion, respectively. Let 
\begin{equation}
 \mathcal{B}=\{\left(u,v\right)\mid B\left(u,v\right)=1\}   
\end{equation}
denote the set of boundary pixels. The distance from each pixel  $(u,v)$ to its nearest boundary pixel is then computed as: 
\begin{equation}
D\left(u,v\right)=\min_{\left(u^{\prime },v^{\prime }\right)\in \mathcal{B}}{\sqrt{{\left(u-u^{\prime }\right)}^{2}+{\left(v-v^{\prime }\right)}^{2}}},
\end{equation}

For the i-th patch region $P_{i}$, the patch-level boundary distance is obtained by averaging the pixel-wise distances:
\begin{equation}
    d_{i}\! =\frac{1}{\left|P_{i}\right|}\sum_{\left(u,v\right)\in P_{i}} D\left(u,v\right).
\end{equation}

Based on $d_{i}$, we define a distance-decaying masking probability:
\begin{equation}
    p_{i}\! =p_{min}\! +\left(p_{max}\! -p_{min}\! \right)exp\left(-\frac{d_{i}}{\tau }\right)\! ,
\end{equation}
where $p_{max}\!$  and $p_{min}$ are the maximum and minimum masking probabilities, and $\tau$ controls the decay rate. The masking indicator $m_{i}$ is then independently sampled from a Bernoulli distribution parameterized by $p_{i}$:

\begin{equation}
m_{i}\sim Bernoulli\left(p_{i}\right).
\end{equation}
where $m_{i}=1$ indicates that the i-th label patch is masked for reconstruction, whereas $m_{i}=0$ indicates that it remains visible. 

For partially masked reconstruction, the same boundary-centric sampling mechanism is independently applied to both reference and query labels. Given a training pair $\left(I_{ref},Y_{ref}\right)$ and $\left(I_{q},Y_{q}\right)$, we construct
\begin{equation}
    Z_{ref}^{\left(0\right)}=E\left(Y_{ref},m_{ref}\right), Z_{q}^{\left(0\right)}=E\left(Y_{q},m_{q}\right).
\end{equation}

Here, $E\left(Y,m\right)$ denotes the masked-label embedding operator that applies the patch-wise token rule above to every patch of $Y$ according to $m$. The resulting $Z_{ref}^{\left(0\right)}$ and $Z_{q}^{\left(0\right)}$ are the initial label-token sequences for the reference and query branches, respectively. By repeatedly emphasizing structurally uncertain regions across heterogeneous training targets, this masking strategy biases the shared reconstruction process toward target-agnostic structural recovery rather than appearance-specific patterns.

\subsection{Multi-level Bidirectional Reference-Query Interaction}

Building on the boundary-focused structural learning, the next step is to identify the reference-defined target in the query despite their appearance discrepancy. To this end, US-Seg establishes multi-level bidirectional interaction between the reference and query branches, allowing target-specific information from the annotated reference to be aligned with the query at multiple semantic depths. Given the joint input sequences, a shared ViT-L encoder extracts reference and query features from a set of intermediate layers $L$, denoted as ${\{x_{ref}^{l}\}}_{l\in \mathcal{L}}\!$ and $\ {\{x_{q}^{l}\}}_{l\in \mathcal{L}}$. These features provide complementary information across semantic depths, with shallower layers preserving local boundary and texture cues and deeper layers encoding higher-level structural semantics.

For each level $l$, bidirectional cross-attention is applied between the two branch representations. The query branch attends to the reference to retrieve target-specific structural cues, while the reference branch attends to the query image representation to contextualize these cues with respect to the current query appearance:
\begin{equation}
\widetilde{x_{q}^{l}}=x_{q}^{l}+\text{CA}\left(x_{q}^{l}, x_{ref}^{l}\right), \widetilde{x_{ref}^{l}}=x_{ref}^{l}+\text{CA}\left(x_{ref}^{l}, x_{q}^{l}\right).
\end{equation}

Here, $CA(\cdot)$ denotes multi-head cross-attention. This reciprocal interaction establishes target-aware reference-query correspondence before feature fusion, rather than relying on one-way reference injection.

The interacted features are then adaptively fused using a learnable spatial gate. Specifically, the query feature, reference feature, and their difference are concatenated to predict the fusion gate:
\begin{equation}
    g^{l}=\mathrm{\sigma }\left(MLP\left(\left[\widetilde{x_{q}^{l}};\widetilde{x_{ref}^{l}};\widetilde{x_{q}^{l}}\! -\widetilde{x_{ref}^{l}}\! \right]\right)\right).
\end{equation}

The fused feature at level l is obtained by
\begin{equation}
f^{l}=g^{l}\odot \! \widetilde{x_{q}^{l}}+\left(1-g^{l}\right)\odot \widetilde{x_{ref}^{l}}.
\end{equation}

The gate dynamically balances query-specific visual evidence and reference-provided structural guidance at each spatial location. Finally, the fused representations from all selected levels are concatenated along the channel dimension:
\begin{equation}
M=\text{Concat}\left(f^{l_{1}},f^{l_{2}},\ldots ,f^{l_{K}}\right).
\end{equation}

The resulting representation integrates reference-query correspondence across multiple semantic depths and serves as the conditioned feature field for subsequent iterative refinement.

\subsection{Prediction-Aware Iterative Refinement}
Given the reference-conditioned multi-level representation, US-Seg employs a prediction-aware iterative refinement decoder for progressive label reconstruction. Instead of producing the query mask through one-shot decoding, the decoder maintains an evolving latent state and repeatedly feeds its intermediate prediction back into the refinement process.

Specifically, the decoder initializes a latent state $s_{0}$ and a persistent memory $m$ from the conditioned representation $M$:

\begin{equation}
s_{0}\! =W_{s}\! M,\ \ m=W_{m}\! M.
\end{equation}

The latent state represents the evolving reconstruction hypothesis, while the fixed memory preserves reference-conditioned target evidence throughout refinement, providing a stable structural guide as the prediction is progressively updated.

At refinement step $t$, the current state $s_{t}$ is decoded into an intermediate mask prediction $\widehat{Y_{t}}$. This prediction is then re-tokenized and projected into the latent space to form a prediction-aware feature $p_{t}$, allowing the current segmentation hypothesis to directly influence the next refinement step. Based on $s_{t}$ and $p_{t}$, a gating module estimates a spatial refinement gate $g_{t}$, which determines where and to what extent the current state should be updated.

The state correction is produced by a shared refinement block $R(\cdot)$, which combines self-attention over the current state with cross-attention to the fixed reference-conditioned memory, while incorporating the prediction-feedback feature $p_{t}$. The state transition is written as:
\begin{equation}
    s_{t+1}\! =s_{t}\! +g_{t}\! \odot R\left(s_{t}\! ,m,p_{t}\! \right).
\end{equation}

Here, $g_t$ determines where and how strongly refinement is needed, while $R(\cdot)$ corrects these regions using prediction feedback and reference-conditioned memory. Importantly, refinement updates only the latent state rather than the model parameters, preserving finetuning-free inference. After $T$ refinement steps, the final query mask is decoded from the last latent state $\widehat{Y_{q}}=\text{Dec}\left(s_{T}\right).$ 

Through this closed-loop process, US-Seg converts prediction feedback into reference-guided state correction, progressively resolving local structural ambiguities and residual segmentation errors without target-specific parameter adaptation.

\subsection{Training Objective and Inference}
US-Seg is optimized with two complementary objectives, including an iterative reconstruction loss that supervises label recovery throughout the refinement process, and a gate regularization term that encourages localized state updates. The overall training objective is:
\begin{equation}
   {\mathcal{L}}_{total}={\mathcal{L}}_{iter}+{\mathrm{\lambda }}_{g}{\mathcal{L}}_{gate}, 
\end{equation}
where ${\mathcal{L}}_{iter}$ supervises reconstruction accuracy across refinement steps and ${\mathcal{L}}_{gate}$ regularizes the spatial refinement gates.

To support both structure-aware learning and the fully masked inference setting, US-Seg adopts a mixed masked-label reconstruction strategy. Under partial-mask training, the reference and query labels are independently corrupted using the boundary-centric masking mechanism described in~\ref{sec:blm}, and the model reconstructs the missing label regions in both branches. Under full-query-mask training, all query label tokens are masked, requiring the query mask to be reconstructed entirely from the annotated reference and query image. This directly exposes the model to the target-absent query condition encountered during inference.

For each branch $b\in\{ref,q\}$, the reconstruction loss at refinement step t is computed only over the masked regions:
\begin{equation}
\mathcal{L}_{rec,b}^{(t)}=\frac{\sum \mathrm{\ell }\left(\widehat{Y_{b}^{t}},Y_{b}\right)\odot M_{b}}{\sum M_{b}+\mathrm{\epsilon }},
\end{equation}
where $M_{b}$ denotes the pixel-level mask expanded from the corresponding patch-level masking indicators, and $\mathrm{\ell }\left(\cdot\right)$ denotes the pixel-wise Smooth L1 loss with $\beta=0.01$. The reconstruction losses from both branches are aggregated across refinement steps as:
\begin{equation}
    {\mathcal{L}}_{iter}=\sum_{t=1}^{T} {\mathrm{\lambda }}_{t}\left({\mathcal{L}}_{rec,ref}^{\left(t\right)}+{\mathcal{L}}_{rec,q}^{\left(t\right)}\right),
\end{equation}
where ${\mathrm{\lambda }}_{t}$ controls the contribution of each refinement step.
To prevent unnecessary global modification of the latent state, we additionally regularize the spatial refinement gate. Let $g_{t}$ denote the refinement gate at step t. The gate regularization is defined as
\begin{equation}
    {\mathcal{L}}_{gate}=\sum_{t=1}^{T} \text{mean}\left(g_{t}\right).
\end{equation}

Minimizing this term encourages sparse and localized updates, so that refinement is concentrated on regions requiring structural correction rather than uniformly modifying the entire prediction. 

During inference, only the annotated reference pair $\left(I_{ref},Y_{ref}\right)$ and query image $I_{q}$ are provided, while all query-side label tokens are initialized as mask tokens. The annotated reference specifies the target, and the decoder progressively reconstructs and refines the query mask through prediction feedback and reference-conditioned guidance.

\section{Experiments}
\subsection{Datasets}
\label{dataset}

\begin{table*}[t]
\centering
\caption{Summary of the ultrasound datasets used in our experiments.}
\label{tab:datasets}
\begin{tabularx}{0.99\textwidth}{lllll}
\toprule
\textbf{Setting} &
\textbf{Organs} &
\textbf{Targets} &
\textbf{Datasets} &
\textbf{\# Images} \\
\midrule

\multirow[c]{7}{2.0cm}{\centering\textbf{In-domain}}
& Breast
& Breast tumor
& BrEaST~\cite{breast}, BUS-BRA~\cite{busbra}, BUSI~\cite{busi}, BUSI-WHU~\cite{busi-whu}
& 3,838 \\

& Thyroid
& Thyroid nodule and thyroid gland
& DDTI~\cite{ddti}, TG3K~\cite{tn3ktg3k}, TN3K~\cite{tn3ktg3k}
& 7,715 \\

& Heart
& LV cavity, myocardium, and left atrium
& CAMUS~\cite{camus}
& 2,000 \\

& Lumbar
& Multifidus muscle
& LUMINOUS~\cite{luminous}
& 341 \\

& Neck
& Brachial plexus
& Neck~\cite{neck}
& 5,635 \\

& Pelvis
& Pubic symphysis and fetal head
& PSFHS~\cite{psfhs}
& 1,358 \\

& Stomach
& Gastrointestinal stromal tumor
& USD514~\cite{usd514}
& 514 \\
\midrule

\multirow[c]{2}{2.0cm}{\centering
\textbf{Out-of-domain}\\
\textbf{Cross-site}}
& Breast
& Breast tumor
& BUS-UCLM~\cite{bus-uclm}, UDIAT~\cite{udiat}
& 427 \\

& Thyroid
& Thyroid nodule
& TNSCUI~\cite{tnscui}
& 3,644 \\
\midrule

\multirow[c]{4}{2.0cm}{\centering
\textbf{Out-of-domain}\\
\textbf{Cross-task}}
& Kidney
& Kidney
& KidneyUS~\cite{kidneyus}
& 487 \\

& Cervix
& Anterior and posterior cervical lips
& FUGC~\cite{fugc}
& 440 \\

& Prostate
& Prostate gland
& MicroSeg~\cite{microseg}
& 2,621 \\

& Ovary
& Ovarian tumor
& MMOTU~\cite{mmotu}
& 1,469 \\
\bottomrule
\end{tabularx}
\end{table*}

We evaluate US-Seg on diverse ultrasound datasets spanning multiple anatomical targets, imaging views, devices, and clinical sources, as summarized in Table I. All splits are performed at the patient level to avoid information leakage. Accordingly, we organize the evaluation into three settings of increasing generalization difficulty: 

(1) In-domain setting: Both the segmentation task and the imaging distribution are represented during training. This setting evaluates US-Seg on training-defined targets under familiar imaging conditions.

(2) Cross-site setting: The segmentation task is represented during training, while the test dataset is entirely excluded. This setting evaluates whether the learned segmentation capability can transfer across changes in clinical sources, imaging devices, acquisition protocols, and image characteristics without target-dataset adaptation.

(3) Cross-task setting: Neither the segmentation task nor its dataset is represented during training. The model must infer the target definition from the annotated reference and transfer its learned structural reconstruction capability to a new segmentation task beyond the training task space.

For the in-domain datasets, cases are split at the patient level into training, validation, and test sets with a ratio of 7:1:2, and US-Seg is jointly trained on all training splits. In-domain references are drawn from the training split, whereas out-of-domain references are selected from the corresponding dataset and excluded from the query set, with all remaining out-of-domain (OOD) cases used as queries. The query sets are fixed across all experiments.

During training, reference-query pairs consist of two images from different patients with the same segmentation target. During standard evaluation, US-Seg, PerSAM, and INSID3 use one fixed reference, whereas UniverSeg and Tyche use 16 fixed support examples. For the robustness analysis in~\ref{sec:reference}, the query set remains fixed, and each of the 100 annotated cases is used individually as the reference for evaluation.

Together, these settings provide a progressive evaluation from in-domain segmentation to cross-site transfer and ultimately cross-task generalization.

\subsection{Implementation Details}
US-Seg is implemented in PyTorch and trained on a single NVIDIA A100 GPU. All images and masks are resized to 448×448. We use the Adam optimizer with an initial learning rate of $1\times 10^{-3}$ and a batch size of 32, and train the model for 400 epochs. During training, partial-mask and full-query-mask reconstruction are sampled with probabilities of 0.8 and 0.2, respectively. For boundary-centric masking, we set $p_{\min}=0.4$, $p_{\max}=0.8$, and $\tau=24$ pixels. The iterative decoder performs T=3 refinement steps with $\left({\mathrm{\lambda }}_{1},{\mathrm{\lambda }}_{2},{\mathrm{\lambda }}_{3}\right)=\left(0.3,0.7,1.0\right)$, and the gate regularization weight ${\mathrm{\lambda }}_{g}$ is set to 0.01. For each segmentation target, a single annotated reference image-mask pair from a different patient is used to specify the target and can be reused across multiple query images. 

\subsection{Compared Methods and Evaluation Metrics}

We compare US-Seg with representative methods from four segmentation paradigms, including finetuning-based methods (nnUNet~\cite{nnunet} and USFM~\cite{usfm}), spatial-prompted methods (MedSAM~\cite{medsam}, UltraSAM~\cite{ultrasam}, and SAMUS~\cite{samus}), task-prompted ultrasound segmentation methods (MOFO~\cite{mofo} and PerceptGuide~\cite{perceptguide}), and label-reference methods (UniverSeg~\cite{universeg}, Tyche~\cite{tyche}, PerSAM~\cite{C2}, and INSID3~\cite{insid3}). For spatial-prompted methods, a single point prompt is provided for each query image. Following their respective learning paradigms, nnUNet and USFM are independently finetuned on each in-domain dataset. For UltraSAM, SAMUS, MOFO, and PerceptGuide, whose original training data overlap with part of our evaluation datasets, we retrain the models using only our in-domain training data to avoid information leakage. For MedSAM, UniverSeg, Tyche, PerSAM, and INSID3, we directly use their publicly released pretrained weights without additional retraining or adaptation, as these models have already been trained on large-scale medical image datasets and do not involve overlap with our evaluation sets. 

For label-reference methods, we follow their respective reference paradigms. US-Seg, PerSAM, and INSID3 use a single annotated reference, whereas UniverSeg and Tyche are evaluated with 16 annotated support examples. Segmentation performance is evaluated using Dice and intersection over union (IoU).

\subsection{Universal Segmentation across Ultrasound Scenarios}

\subsubsection{Effectiveness of Unified In-domain Learning}

\begin{table*}[t]
\centering
\caption{Comparison of Different Models on In-domain Datasets. * indicates evaluation with 16 annotated references.}
\label{tab:id_results}
\scriptsize
\resizebox{\textwidth}{!}{
\begin{tabular}{l*{14}{c}}
\toprule
\multirow{2}{*}{\textbf{Method}}
& \multicolumn{2}{c}{\textbf{Breast}}
& \multicolumn{2}{c}{\textbf{Thyroid}}
& \multicolumn{2}{c}{\textbf{Heart}}
& \multicolumn{2}{c}{\textbf{Lumbar}}
& \multicolumn{2}{c}{\textbf{Neck}}
& \multicolumn{2}{c}{\textbf{Pelvis}}
& \multicolumn{2}{c}{\textbf{Stomach}} \\
\cmidrule(lr){2-3}
\cmidrule(lr){4-5}
\cmidrule(lr){6-7}
\cmidrule(lr){8-9}
\cmidrule(lr){10-11}
\cmidrule(lr){12-13}
\cmidrule(lr){14-15}
& \textbf{Dice} & \textbf{IoU}
& \textbf{Dice} & \textbf{IoU}
& \textbf{Dice} & \textbf{IoU}
& \textbf{Dice} & \textbf{IoU}
& \textbf{Dice} & \textbf{IoU}
& \textbf{Dice} & \textbf{IoU}
& \textbf{Dice} & \textbf{IoU} \\
\midrule

\multicolumn{15}{l}{\textcolor{gray}{\itshape Finetuning-based methods}} \\
nnUNet 
& 86.28 & 80.36 
& 79.37 & 71.82 
& 90.25 & 82.17 
& 91.14 & 84.69 
& 80.10 & 69.80 
& 96.44 & 93.02 
& 79.77 & 69.91 \\
USFM 
& 87.16 & 80.62 
& 80.69 & 72.14 
& 90.46 & 82.71 
& 92.04 & 84.84
& 79.25 & 68.32  
& 97.36 & 94.73 
& 81.87 & 71.09 \\
\midrule

\multicolumn{15}{l}{\textcolor{gray}{\itshape Spatial-prompted methods}} \\
% SAM3
% & 63.18 & 50.92 
% & 58.74 & 43.86 
% & 60.35 & 49.71 
% & 82.16 & 72.94 
% & 38.69 & 25.77 
% & 55.42 & 42.08 
% & 24.36 & 16.82 \\
MedSAM
& 63.58 & 52.22 
& 65.43 & 52.77 
& 49.05 & 41.24 
& 56.01 & 41.10 
& 76.22 & 63.77 
& 73.16 & 63.71 
& 46.29 & 33.59 \\
UltraSAM
& 75.77 & 67.82
& 76.63 & 68.29
& 90.06 & 81.58
& 86.40 & 81.93
& 69.84 & 57.36
& 91.61 & 84.86
& 74.39 & 64.02 \\
SAMUS
& 87.24 & 80.07 
& 81.13 & 74.71 
& 90.60 & 83.26 
& 90.45 & 83.99 
& 80.95 & 69.32
& 94.98 & 91.35 
& 81.69 & 72.48 \\
\midrule
\multicolumn{15}{l}{\textcolor{gray}{\itshape Task-prompted methods}} \\
MOFO
& 83.84 & 76.01
& 76.45 & 67.91
& 89.76 & 80.05
& 88.95 & 82.34
& 73.44 & 61.02
& 94.02 & 88.91
& 80.58 & 70.21 \\
PerceptGuide
& 83.23 & 75.48
& 79.23 & 71.54
& 88.12 & 79.67
& 87.51 & 81.10
& 78.71 & 67.18
& 93.27 & 87.50
& 76.24 & 65.73 \\
\midrule
\multicolumn{15}{l}{\textcolor{gray}{\itshape Label-reference methods}} \\
PerSAM
& 22.52 & 17.14 
& 23.11 & 13.66 
& 29.45 & 15.40 
& 27.18 & 16.05 
&  5.73 &  2.96 
& 24.78 & 14.26 
&  8.19 &  4.47 \\
INSID3 
& 28.52 & 21.41 
& 24.77 & 16.10 
& 25.89 & 14.75 
& 24.28 & 13.94 
& 7.76  & 4.09 
& 23.31 & 14.52 
& 13.66 & 7.85  \\
UniverSeg* 
& 56.65 & 44.37 
& 36.60 & 25.59 
& 52.17 & 37.93 
& 82.87 & 72.51 
& 30.25 & 20.39 
& 48.90 & 35.26 
& 27.59 & 19.50 \\
Tyche*
& 57.91 & 45.15 
& 36.90 & 25.92 
& 52.95 & 38.85 
& 77.14 & 64.21 
& 25.12 & 16.82
& 42.87 & 30.47 
& 18.41 & 13.08 \\
\textbf{US-Seg} 
& \textbf{88.48} & \textbf{81.44} 
& \textbf{83.23} & \textbf{76.03} 
& \textbf{91.04} & \textbf{83.90} 
& \textbf{92.89} & \textbf{85.48} 
& \textbf{81.10} & \textbf{70.12} 
& \textbf{97.54} & \textbf{94.75}
& \textbf{83.32} & \textbf{74.49} \\

\bottomrule
\end{tabular}
}
\end{table*}

\begin{table*}[t]
\centering
\caption{Comparison of Different Models on Out-of-Domain Datasets. * indicates evaluation with 16 annotated references.}
\label{tab:ood_comparison}
\renewcommand{\arraystretch}{1}
\setlength{\tabcolsep}{3.2pt}
\scriptsize
\resizebox{\textwidth}{!}{%
\begin{tabular}{ll*{18}{c}}
\toprule
\multirow{3}{*}{\textbf{Type}} 
& \multirow{3}{*}{\textbf{Organ}} 
& \multicolumn{4}{c}{\textbf{Finetuning-based}} 
& \multicolumn{4}{c}{\textbf{Spatial-prompted}} 
& \multicolumn{4}{c}{\textbf{Task-prompted}} 
& \multicolumn{6}{c}{\textbf{Label-reference}}  \\
\cmidrule(lr){3-6}
\cmidrule(lr){7-10}
\cmidrule(lr){11-14}
\cmidrule(lr){15-20}
&
& \multicolumn{2}{c}{\textbf{nnUNet}} 
& \multicolumn{2}{c}{\textbf{USFM}} 
& \multicolumn{2}{c}{\textbf{MedSAM}} 
& \multicolumn{2}{c}{\textbf{SAMUS}} 
& \multicolumn{2}{c}{\textbf{MOFO}} 
& \multicolumn{2}{c}{\textbf{PerceptGuide}} 
& \multicolumn{2}{c}{\textbf{UniverSeg*}} 
& \multicolumn{2}{c}{\textbf{Tyche*}} 
& \multicolumn{2}{c}{\textbf{US-Seg}} \\
\cmidrule(lr){3-4}
\cmidrule(lr){5-6}
\cmidrule(lr){7-8}
\cmidrule(lr){9-10}
\cmidrule(lr){11-12}
\cmidrule(lr){13-14}
\cmidrule(lr){15-16}
\cmidrule(lr){17-18}
\cmidrule(lr){19-20}
&
& \textbf{Dice} & \textbf{IoU}
& \textbf{Dice} & \textbf{IoU}
& \textbf{Dice} & \textbf{IoU}
& \textbf{Dice} & \textbf{IoU}
& \textbf{Dice} & \textbf{IoU}
& \textbf{Dice} & \textbf{IoU}
& \textbf{Dice} & \textbf{IoU}
& \textbf{Dice} & \textbf{IoU}
& \textbf{Dice} & \textbf{IoU} \\
\midrule

\multirow{2}{*}{Cross-site}
& Breast
& 61.68 & 53.17
& 70.30 & 62.06
& 62.11 & 51.20
& 83.93 & 75.20
& 78.59 & 71.72
& 76.47 & 67.89
& 69.84 & 58.56
& 57.45 & 46.52
& \textbf{86.24} & \textbf{78.98} \\

& Thyroid
& 61.49 & 51.81
& 68.01 & 58.27
& 73.96 & 62.77
& 75.96 & 68.67
& 66.14 & 55.81
& 72.54 & 63.20
& 37.39 & 26.43
& 38.02 & 26.53
& \textbf{80.81} & \textbf{72.40} \\

\midrule

\multirow{4}{*}{Cross-task}
& Kidney
& --    & --
& --    & --
& 64.43 & 50.03
& 78.89 & 67.24
& --    & --
& --    & --
& 69.96 & 58.12
& 65.11 & 51.03
& \textbf{82.95} & \textbf{72.81} \\

& Cervix
& --    & --
& --    & --
& 46.39 & 36.83
& 75.58 & 67.82
& --    & --
& --    & --
& 53.29 & 40.54
& 51.25 & 38.37
& \textbf{86.04} & \textbf{82.48} \\

& Prostate
& --    & --
& --    & --
& 61.83 & 47.22
& 73.51 & 65.05
& --    & --
& --    & --
& 60.33 & 47.42
& 57.71 & 44.15
& \textbf{78.40} & \textbf{69.22} \\

& Ovary
& --    & --
& --    & --
& 66.94 & 54.51
& 70.29 & 63.82
& --    & --
& --    & --
& 54.48 & 41.91
& 57.29 & 44.78
& \textbf{76.68} & \textbf{69.06} \\

\bottomrule
\end{tabular}%
}
\end{table*}

As shown in Table II, US-Seg achieves the best performance across all seven in-domain organ groups, with average Dice and IoU scores of 88.23\% and 80.89\%, respectively. Compared with universal representation learning, USFM provides strong transferable features but still requires dataset-specific adaptation for segmentation. US-Seg surpasses USFM by 1.25\% in Dice and 1.68\% in IoU while using a single model jointly trained across heterogeneous datasets, showing that unified segmentation does not compromise performance under in-domain distributions. Spatial-prompted methods, including MedSAM, UltraSAM, and SAMUS, further improve target flexibility, but they require repeated interaction for individual query images. US-Seg instead uses one annotated reference that can be reused across multiple queries while achieving stronger segmentation performance.

Task-prompted methods such as MOFO and PerceptGuide avoid repeated spatial prompting by specifying targets at the task level, but their predictions remain restricted to targets defined during training. US-Seg improves Dice by 4.37\% over MOFO and IoU by 5.43\% over PerceptGuide, demonstrating that reference-based specification can provide accurate segmentation without relying on a fixed target space. Label-reference methods provide the closest comparison because they also define the target through annotated examples. Despite using publicly released models pretrained on large-scale medical image datasets, and providing UniverSeg and Tyche with 16 annotated support examples, their performance remains substantially lower on heterogeneous ultrasound data. US-Seg outperforms the strongest label-reference baseline, UniverSeg, by 40.37\% in Dice and 44.38\% in IoU. These results indicate that strong pretraining and additional references alone are insufficient when structural generalization and reference-query correspondence remain challenging.

Overall, US-Seg combines strong in-domain performance with reliable label-reference segmentation and reusable target specification, while avoiding repeated spatial interaction and predefined task spaces.

\begin{figure*}[htbp]
    \centering
    \includegraphics[width=0.85\linewidth]{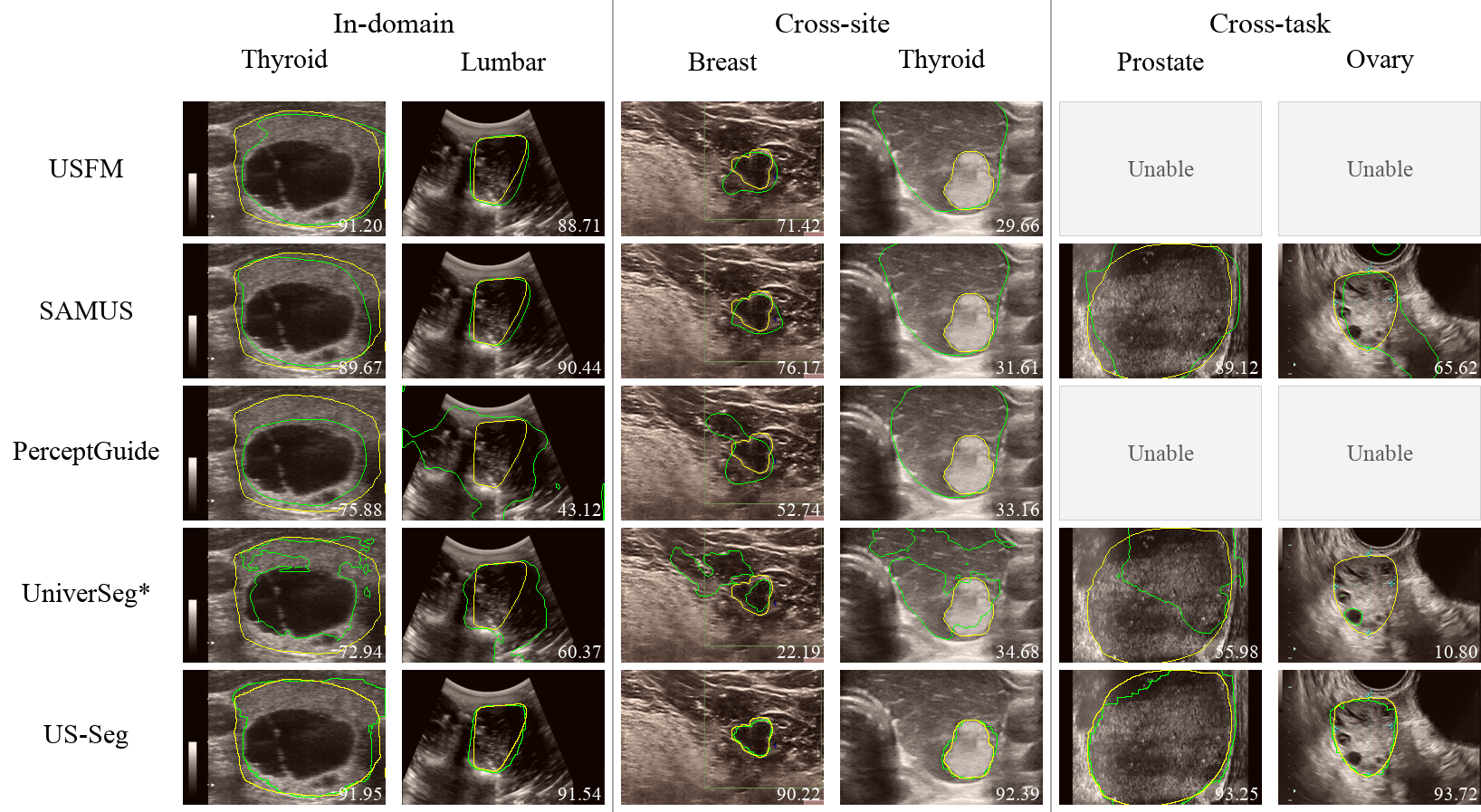}
    \caption{Qualitative comparison across diverse ultrasound segmentation scenarios. Depicted in yellow is the groundtruth, and in green is the predicted result. * indicates evaluation with 16 annotated references.}
    \label{fig3}
\end{figure*}

\subsubsection{Out-of-Domain Generalization across Sites and Tasks}

As reported in Table III, US-Seg achieves the best performance across all six out-of-domain evaluation scenarios, with average Dice and IoU scores of 81.85\% and 74.16\%, respectively. The cross-task setting also exposes a fundamental limitation of finetuning-based and task-prompted methods, with the former requiring target-specific training data and the latter remaining restricted to targets defined during training. In contrast, US-Seg specifies a new target from a single annotated reference and directly performs segmentation without parameter adaptation, enabling the same model to operate across both cross-site and cross-task settings.

Under cross-site evaluation, where the segmentation task is preserved but the imaging distribution changes, US-Seg achieves 83.53\% Dice and 75.69\% IoU. Compared with the strongest spatial-prompted baseline SAMUS, it further improves Dice and IoU by 3.58\% and 3.76\% while replacing repeated query-specific prompting with a single annotated reference that can be reused across queries. US-Seg also substantially outperforms existing label-reference methods, exceeding UniverSeg by 29.92\% in Dice and 33.19\% in IoU, even though UniverSeg and Tyche are provided with 16 annotated support examples. This suggests that increasing the number of references alone is insufficient to overcome the reference-query discrepancy caused by heterogeneous ultrasound appearances. 

The advantage becomes more pronounced under cross-task evaluation, where both the segmentation target and its dataset lie outside the training task space. US-Seg achieves 81.02\% Dice and 73.39\% IoU, while maintaining clear advantages over both spatial-prompted and label-reference methods. Notably, it outperforms the strongest label-reference baseline UniverSeg by 21.50\% in Dice and 26.39\% in IoU, despite using only one reference compared with 16 support examples. These results demonstrate that US-Seg can transfer reference-conditioned structural knowledge across both imaging distributions and segmentation tasks with a single reusable reference.

\subsubsection{Qualitative Superiority across Ultrasound Scenarios}

Fig. 4 presents representative results under in-domain, cross-site, and cross-task settings, with strong methods from each segmentation paradigm selected for comparison. Under in-domain conditions, US-Seg produces accurate and structurally complete predictions across observed targets. Under cross-site shifts, competing methods exhibit boundary deviations, incomplete regions, or false positives, whereas US-Seg remains well aligned with the target structures. This advantage is further pronounced in cross-task cases, where finetuning-based and task-prompted methods are not directly applicable to unseen targets, while US-Seg uses the annotated reference to specify the intended structure and produces more complete predictions than existing spatial-prompted and label-reference methods. These results visually support the robustness of US-Seg across progressively challenging ultrasound scenarios.

\subsection{Target-Specification Efficiency}
\begin{figure}[htbp]
    \centering
    \includegraphics[width=0.95\linewidth]{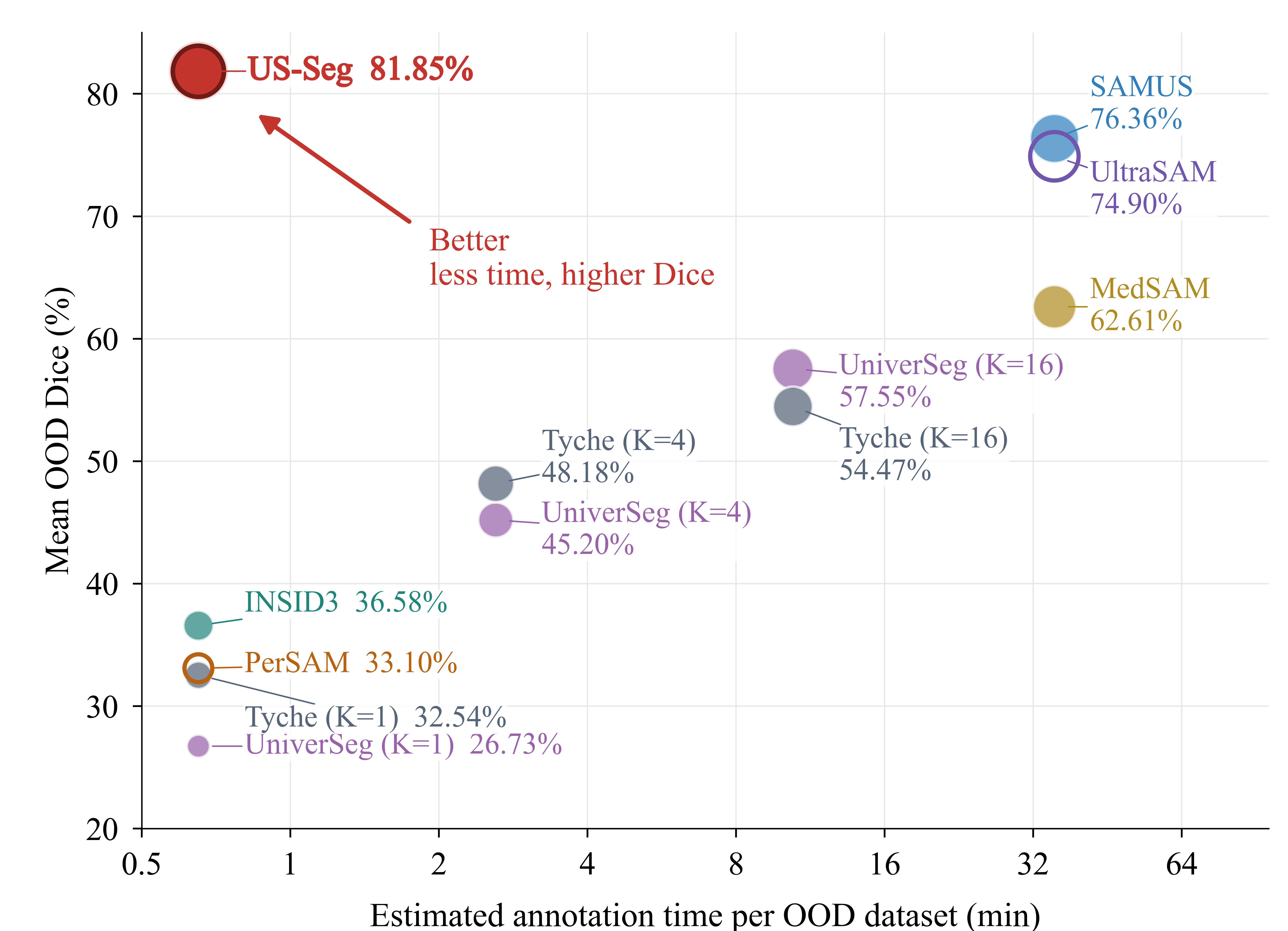}
    \caption{Segmentation Accuracy versus Target-Specification Cost. K indicates evaluation with K annotated references.}
    \label{fig4}
\end{figure}

\begin{figure*}[!t]
    \centering
    \subfloat[Reference-controlled target switching.]{%
        \includegraphics[width=0.55\textwidth]{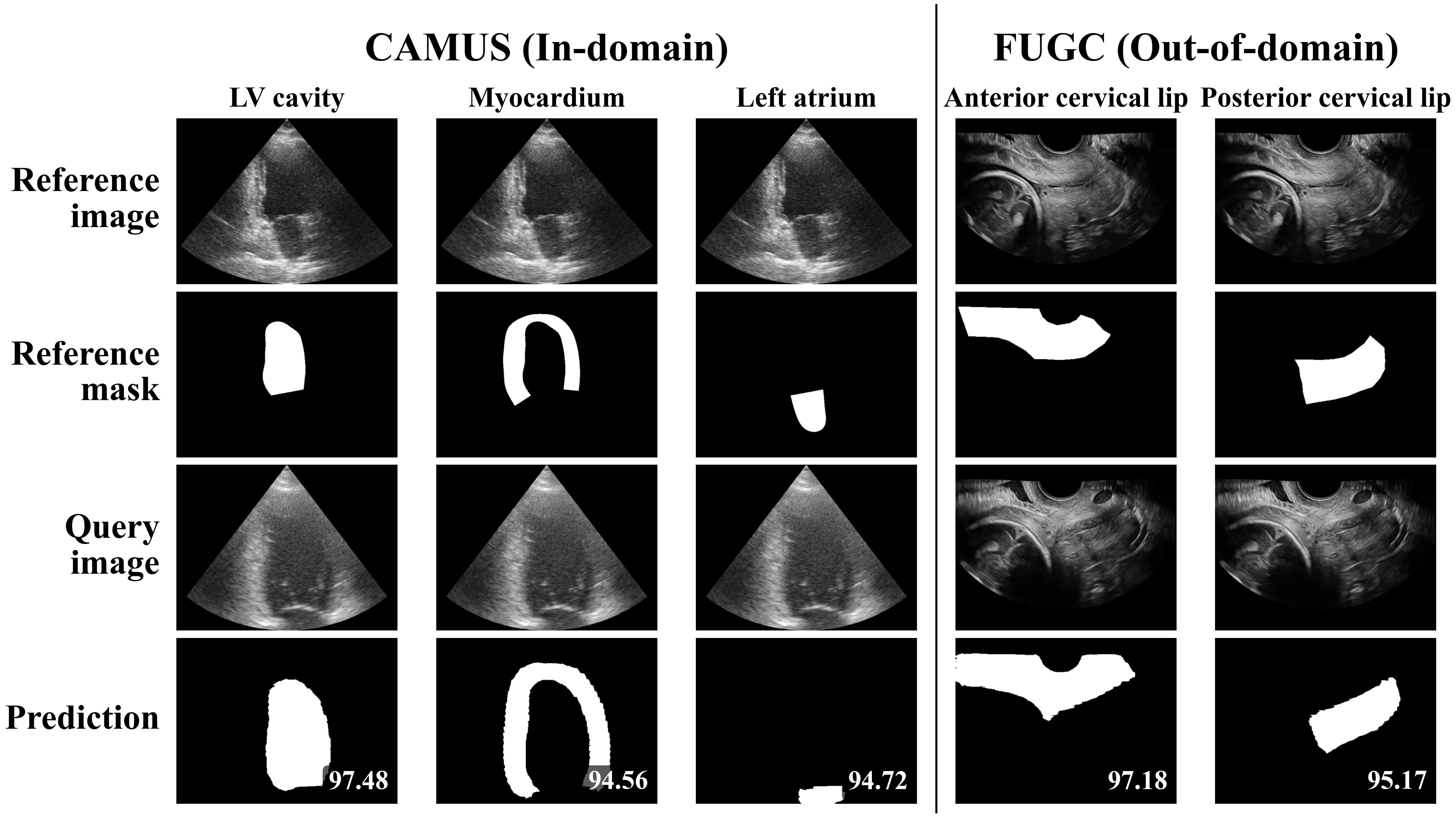}%
        \label{D1}}
    \hfill
    \subfloat[Stability across reference instances.]{%
        \includegraphics[width=0.38\textwidth]{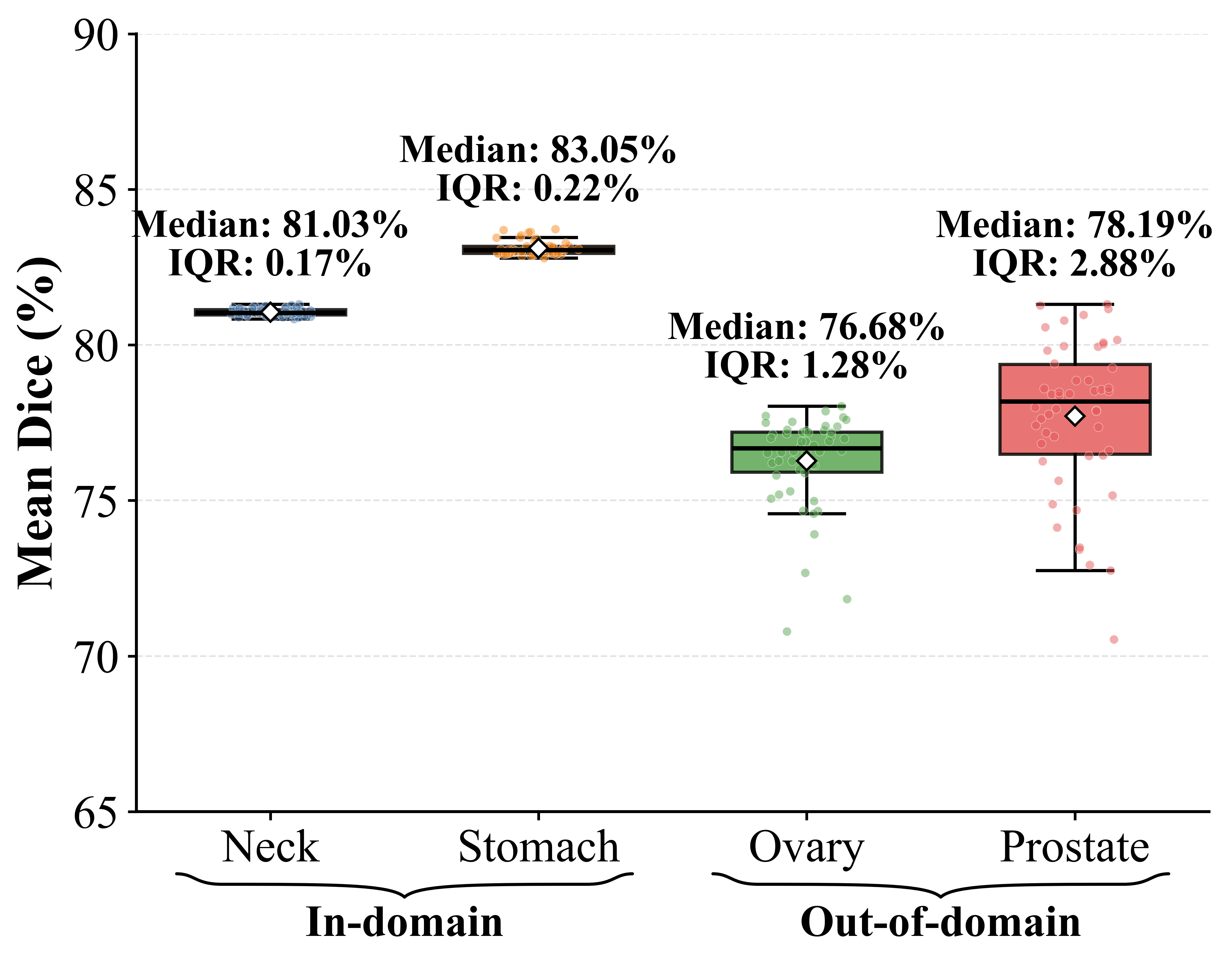}%
        \label{D2}}
    \caption{Reference controllability and robustness of US-Seg.}
    \label{D}
\end{figure*}

Beyond segmentation accuracy, we further evaluate the human effort required for target specification across the six OOD tasks. To compare different interaction paradigms on a common scale, we estimate the total annotation cost required to segment an entire dataset. Annotating a complete reference mask requires 34.2 s on average. 

As shown in Fig. 5, US-Seg achieves the highest mean OOD Dice of 81.85\% with a single annotated reference. In contrast, spatial-prompted methods require target specification separately for every query image, causing their total interaction cost to increase with dataset size. Placing a point prompt requires 1.7 s per query on average, resulting in an estimated annotation cost of approximately 35.2 min per OOD dataset, about 62 times that of US-Seg. Despite this substantially higher interaction cost, SAMUS, UltraSAM, and MedSAM achieve lower Dice scores of 76.36\%, 74.90\%, and 62.61\%, respectively.

Increasing the reference budget improves UniverSeg and Tyche, but using 16 references requires approximately 9.12 min, 16 times the annotation cost of US-Seg, while their Dice scores remain 24.30\% and 27.38\% lower, respectively. PerSAM and INSID3 use the same one-reference budget as US-Seg but achieve substantially lower accuracy, indicating that low annotation cost alone is insufficient without effective structural generalization and reference-query transfer.

Overall, US-Seg provides a more favorable accuracy–effort trade-off by converting target specification from repeated or multi-example annotation into a single reusable reference. This property ensures that its annotation cost does not increase with the number of subsequent queries, supporting scalable deployment across clinical cohorts and external datasets.

\subsection{Reference-Controlled Target Specification and Robustness}
\label{sec:reference}

To further examine whether a single annotated reference can effectively specify the intended target in universal ultrasound segmentation, we first evaluate the target-switching capability of US-Seg by fixing the query image and changing only the reference image-mask pair. As shown in Fig. 6(a), for the same CAMUS query image, US-Seg can accurately switch among the left ventricular cavity, myocardium, and left atrium, achieving Dice scores of 97.48\%, 94.56\%, and 94.72\%, respectively. A similar phenomenon is observed on the out-of-domain FUGC dataset, where the model successfully distinguishes the anterior cervical lip from the posterior cervical lip and achieves Dice scores of 97.18\% and 95.17\%. Since the query image and model parameters remain unchanged, these results demonstrate that the segmentation target is determined by the annotated reference rather than by a fixed output category or the most visually salient structure in the query. Specifically, the reference mask specifies the target identity, while the reference image provides its structural, visual, and anatomical context. US-Seg can therefore switch the intended target solely through the reference without retraining or task-specific redesign.

A practical label-reference model should not only respond correctly to changes in target semantics, but also maintain reliable performance across different reference instances of the same target. Following the evaluation protocol described in Sec.~\ref{dataset}, we use each of the 100 annotated cases individually as the reference and evaluate it on the fixed query set, with the resulting performance distributions summarized in Fig.~6(b). On the in-domain Neck and Stomach datasets, US-Seg exhibits highly concentrated performance distributions, achieving median Dice scores of 81.03\% and 83.05\% with interquartile ranges of only 0.17\% and 0.22\%, respectively. More importantly, this robustness is well preserved under out-of-domain evaluation. On the unseen Ovary and Prostate datasets, the median Dice remains as high as 76.68\% and 78.19\%, while the interquartile ranges are limited to 1.28\% and 2.88\%. The consistently high medians and compact distributions demonstrate that US-Seg can reliably transfer target information from diverse reference instances, even for anatomical structures and data distributions not encountered during training. Consequently, strong segmentation performance can be obtained without carefully selecting or screening a favorable reference case.

\subsection{Progressive Structural Correction with Iterative Refinement}

\begin{figure*}[!t]
    \centering
    \subfloat[Step-wise structural refinement.]{%
        \includegraphics[width=0.6\textwidth]{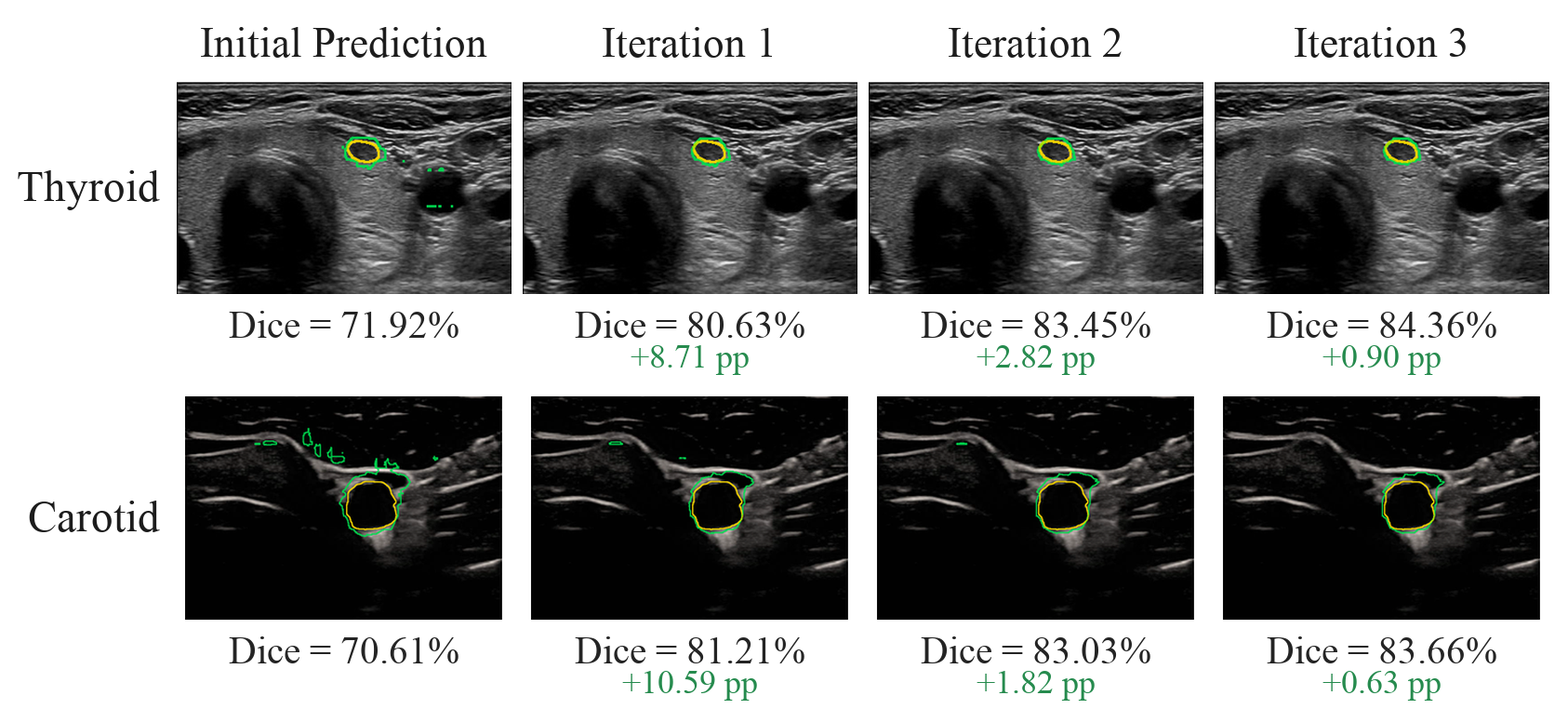}%
        \label{E1}}
    \hfill
    \subfloat[Performance evolution across refinement steps.]{%
        \includegraphics[width=0.38\textwidth]{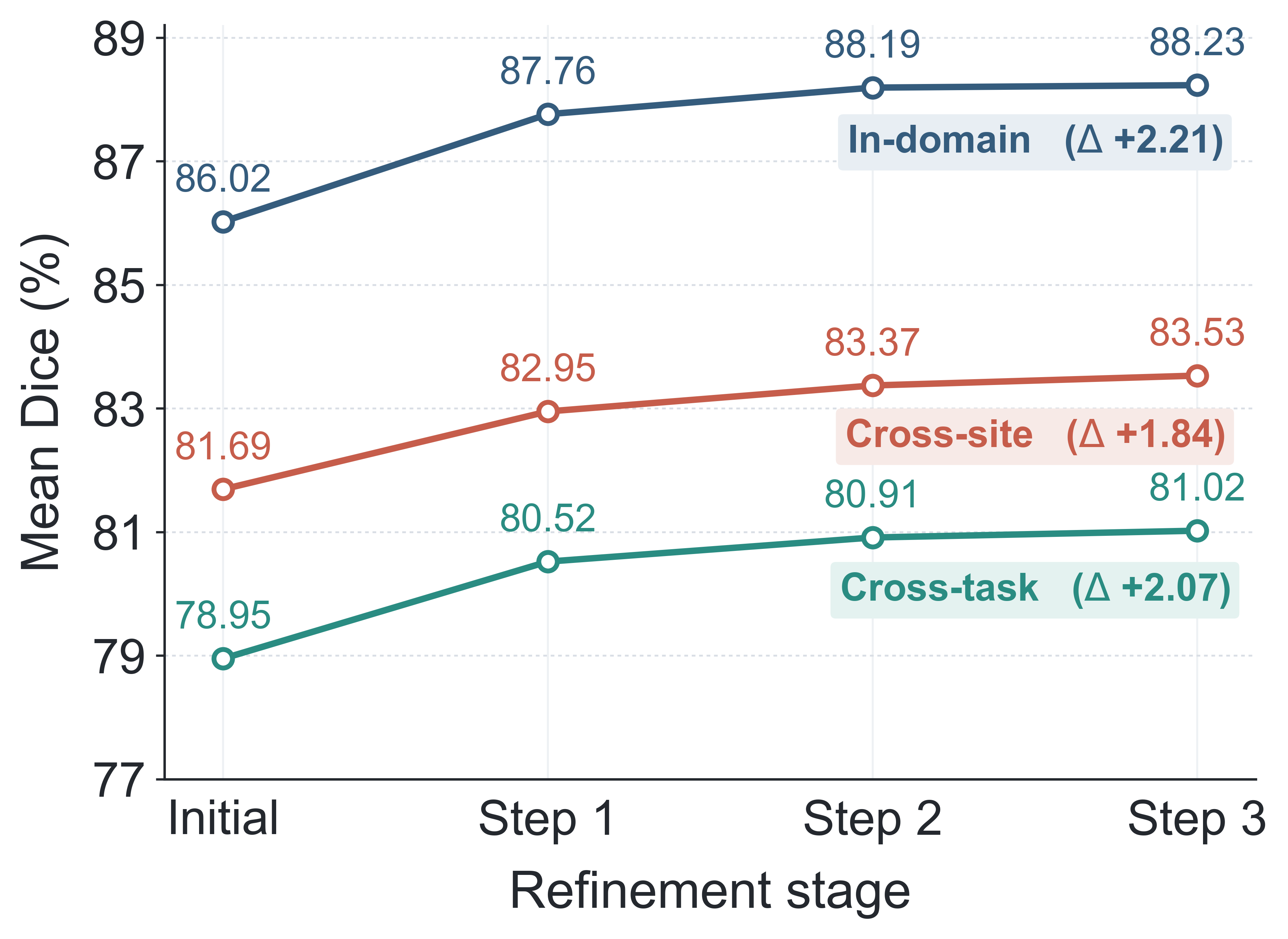}%
        \label{E2}}
    \caption{Progressive structural correction with prediction-aware iterative refinement.  Depicted in yellow is the groundtruth, and in green is the predicted result.}
    \label{D}
\end{figure*}

To examine how prediction-aware iterative refinement corrects residual structural errors, we visualize the evolution of intermediate predictions over three refinement steps. As shown in Fig. 7(a), the initial predictions already localize the target but exhibit evident boundary deviations and incomplete structural recovery. With prediction feedback repeatedly incorporated into the latent-state update, these errors are progressively corrected. For the thyroid example, Dice increases from 71.92\% to 80.63\%, 83.45\%, and 84.36\% over the three refinement steps, with the predicted contour becoming progressively better aligned with the target boundary. A similar trend is observed for the carotid example, where Dice improves from 70.61\% to 81.21\%, 83.03\%, and 83.66\%. These step-wise changes show that the decoder does not simply repeat mask generation; instead, each intermediate prediction guides the subsequent correction of residual structural errors.

We further quantify this behavior over the complete evaluation sets in Fig. 7(b). Iterative refinement increases the mean Dice from 86.02\% to 88.23\% for in-domain evaluation, from 81.69\% to 83.53\% for cross-site evaluation, and from 78.95\% to 81.02\% for cross-task evaluation, corresponding to overall gains of 2.21\%, 1.84\%, and 2.07\%, respectively. Most of the improvement is achieved within the first two refinement steps, where Dice already rises to 88.19\%, 83.37\%, and 80.91\% in the three settings, while the third step contributes only marginal gains of 0.04\%, 0.16\%, and 0.11\%. These diminishing gains indicate progressive convergence while confirming that iterative refinement effectively corrects residual structural errors without parameter updates.

\subsection{Effectiveness of the US-Seg Design}

\begin{table}[t]
\centering
\caption{Ablation study of US-Seg. BLM, MBI,
and PIR denote boundary-centric label masking, multi-level bidirectional reference-query interaction, and prediction-aware iterative refinement.}
\label{tab:component_ablation}
\resizebox{0.48\textwidth}{!}{%
\begin{tabular}{ccc cccccc}
\toprule
\multicolumn{3}{c}{\textbf{Components}}
& \multicolumn{2}{c}{\textbf{In-domain}}
& \multicolumn{2}{c}{\textbf{Cross-site}}
& \multicolumn{2}{c}{\textbf{Cross-task}} \\
\cmidrule(lr){1-3}
\cmidrule(lr){4-5}
\cmidrule(lr){6-7}
\cmidrule(lr){8-9}
\textbf{BLM} & \textbf{MBI} & \textbf{PIR}
& \textbf{Dice} & \textbf{IoU}
& \textbf{Dice} & \textbf{IoU}
& \textbf{Dice} & \textbf{IoU} \\
\midrule
$\checkmark$ &              &
& 82.21 & 74.80
& 72.34 & 63.88
& 69.72 & 57.96 \\

$\checkmark$ &              & $\checkmark$
& 84.67 & 76.85
& 76.91 & 67.25
& 72.34 & 60.89 \\

             & $\checkmark$ & $\checkmark$
& 85.48 & 76.39
& 76.35 & 66.41
& 73.92 & 62.31 \\

$\checkmark$ & $\checkmark$ &
& 86.02 & 77.31
& 81.69 & 73.05
& 78.95 & 69.90 \\

$\checkmark$ & $\checkmark$ & $\checkmark$
& \textbf{88.23} & \textbf{80.89}
& \textbf{83.53} & \textbf{75.69}
& \textbf{81.02} & \textbf{73.39} \\
\bottomrule
\end{tabular}
}
\end{table}

Table IV evaluates the contributions of boundary-centric label masking (BLM), multi-level bidirectional reference-query interaction (MBI), and prediction-aware iterative refinement (PIR) under in-domain, cross-site, and cross-task settings. The complete US-Seg achieves the best performance in all three settings, with Dice scores of 88.23\%, 83.53\%, and 81.02\%, respectively.

Removing BLM decreases Dice by 2.75\%, 7.18\%, and 7.10\% under the in-domain, cross-site, and cross-task settings, respectively. The substantially larger degradation under out-of-domain evaluation highlights the importance of boundary-centric reconstruction for structural generalization beyond familiar imaging distributions. Removing MBI leads to Dice drops of 3.56\%, 6.62\%, and 8.68\%, with the largest decrease observed under cross-task evaluation. The corresponding cross-task IoU also decreases by 12.50\%. This result confirms the importance of multi-level bidirectional interaction for transferring the reference-defined target across reference-query discrepancies. Removing PIR results in consistent Dice decreases of 2.21\%, 1.84\%, and 2.07\% across the three settings, demonstrating that iterative prediction feedback provides additional gains by correcting residual structural errors.

Overall, BLM, MBI, and PIR provide complementary benefits for structural generalization, reference-guided target transfer, and residual error correction, respectively.

\section{Discussion}

US-Seg advances universal ultrasound segmentation from transferable representation learning toward direct, reference-specified segmentation, eliminating the need for target-specific adaptation. We next discuss how this formulation extends segmentation-level universality, changes the paradigm of target specification, and supports more scalable clinical interaction, together with its remaining limitations.

\subsection{From Universal Representation to Universal Segmentation}

Ultrasound foundation models have demonstrated that heterogeneous pretraining can produce representations transferable across organs, imaging conditions, and downstream tasks. However, this representation-level universality does not directly yield universal segmentation. Existing frameworks still rely on task-specific decoders or target-data adaptation to convert transferable features into segmentation outputs, meaning that their universality primarily resides in feature reuse rather than direct, target-flexible mask generation.

US-Seg extends universality from reusable representations to direct, target-flexible segmentation. By using an annotated reference to define the target at inference, the same model can operate across imaging sites and instantiate segmentation tasks beyond the training task space without modifying its output space or finetuning on target data. Importantly, the jointly trained US-Seg also outperforms the dataset-specific USFM baseline by 1.25\% in average in-domain Dice, indicating that broader generalization does not require sacrificing performance on familiar distributions. These results suggest that universal ultrasound segmentation requires not only transferable features, but also a prediction mechanism that can translate shared structural knowledge into directly executable segmentation tasks.

\subsection{From Prompt-Driven Interaction to Reusable Target Specification}

Existing universal ultrasound segmentation methods mainly achieve target flexibility through task-level or spatial prompting. Task-prompted methods, such as MOFO and PerceptGuide, encode target identity explicitly but remain restricted to segmentation tasks defined during training. Spatial-prompted methods, including MedSAM, UltraSAM, and SAMUS, accommodate more flexible targets but require query-specific interaction, resulting in approximately 62 times the annotation cost of US-Seg in OOD evaluation while remaining 5.49\% lower in Dice. These two paradigms trade predefined target identity for query-specific flexibility, but neither provides a reusable mechanism for target specification across images.

Label-reference methods provide a more reusable form of target specification by defining the intended structure through annotated examples. Such approaches still face a trade-off between annotation burden and transfer reliability. UniverSeg and Tyche improve robustness by using multiple annotated supports, whereas one-reference methods such as PerSAM and INSID3 substantially reduce annotation requirements but remain sensitive to large reference-query discrepancies. In our experiments, even with 16 annotated references, the strongest multi-reference baseline UniverSeg remains 24.30\% lower in OOD Dice than US-Seg, while the strongest one-reference baseline INSID3 remains 45.27\% lower under the same reference budget. These results suggest that effective reference-based segmentation depends not simply on increasing the number of annotations, but on learning transferable structure and establishing robust reference-query correspondence. 

US-Seg addresses this trade-off by combining single-reference target specification with transferable structural learning and robust reference-query interaction, enabling a single annotated reference to consistently define the target across queries without repeated spatial interaction or target-specific adaptation.

\subsection{Clinical Significance of US-Seg}

US-Seg offers a practical way to extend ultrasound segmentation to clinical scenarios where targets, imaging conditions, or datasets may change over time. Instead of requiring a new model, additional finetuning, or repeated per-image prompting whenever a new setting is encountered, clinicians can define the intended target once through an annotated reference and reuse it across subsequent cases. This is particularly relevant for cohort-level analysis, longitudinal follow-up, retrospective database studies, and newly introduced measurements, where repeated interaction or target-specific training would substantially limit scalability.

The main clinical value of this paradigm lies in converting target specification from a repeated per-image operation into a reusable one-time instruction. Although producing a complete reference mask requires more effort than a single point prompt, this cost can be amortized over subsequent queries. Moreover, the observed robustness to different reference instances suggests that deployment does not depend on carefully selecting an exceptional reference case. Together with adaptation-free cross-site and cross-task generalization, these properties make US-Seg particularly suitable for scalable clinical workflows in which segmentation needs evolve without corresponding increases in annotation and model-maintenance burden.

\subsection{Limitations and Future Directions}

The current study focuses primarily on two-dimensional B-mode ultrasound. Future work will extend US-Seg to broader ultrasound settings, including temporal sequences, three-dimensional imaging, and multimodal acquisitions, to exploit richer anatomical and functional information. In addition, although the present evaluation covers diverse organs, clinical sources, and challenging out-of-domain scenarios, further validation on larger prospective multi-center cohorts will be valuable for assessing its applicability across broader clinical populations and workflows.

\section{Conclusion}
This work presents US-Seg, a universal ultrasound segmentation framework that uses a single annotated reference to specify the target and directly segments query images without target-specific adaptation. By formulating heterogeneous segmentation tasks within a unified reference-guided framework, US-Seg enables a single model to maintain strong in-domain performance while generalizing effectively across imaging sites and segmentation tasks. Extensive experiments further demonstrate accurate and stable target specification from only one reusable reference, with substantially lower annotation and interaction costs than repeated spatial prompting or multi-reference support. These findings establish a scalable and generalizable pathway toward universal ultrasound segmentation beyond fixed imaging distributions and predefined task spaces.

\bibliographystyle{ieeetr}
\bibliography{main}

@article{usfm,
  title={Usfm: A universal ultrasound foundation model generalized to tasks and organs towards label efficient image analysis},
  author={Jiao, Jing and Zhou, Jin and Li, Xiaokang and Xia, Menghua and Huang, Yi and Huang, Lihong and Wang, Na and Zhang, Xiaofan and Zhou, Shichong and Wang, Yuanyuan and others},
  journal={Medical image analysis},
  volume={96},
  pages={103202},
  year={2024},
  publisher={Elsevier}
}

@inproceedings{sam,
  title={Segment anything},
  author={Kirillov, Alexander and Mintun, Eric and Ravi, Nikhila and Mao, Hanzi and Rolland, Chloe and Gustafson, Laura and Xiao, Tete and Whitehead, Spencer and Berg, Alexander C and Lo, Wan-Yen and others},
  booktitle={2023 IEEE/CVF international conference on computer vision (ICCV)},
  pages={3992--4003},
  year={2023},
  organization={IEEE}
}

@article{medsam,
  title={Segment anything in medical images},
  author={Ma, Jun and He, Yuting and Li, Feifei and Han, Lin and You, Chenyu and Wang, Bo},
  journal={Nature communications},
  volume={15},
  number={1},
  pages={654},
  year={2024},
  publisher={Nature Publishing Group UK London}
}

@inproceedings{samus,
  title={Beyond adapting SAM: Towards end-to-end ultrasound image segmentation via auto prompting},
  author={Lin, Xian and Xiang, Yangyang and Yu, Li and Yan, Zengqiang},
  booktitle={International conference on medical image computing and computer-assisted intervention},
  pages={24--34},
  year={2024},
  organization={Springer}
}

@article{ultrasam,
  title={Ultrasam: a foundation model for ultrasound using large open-access segmentation datasets},
  author={Meyer, Adrien and Murali, Aditya and Zarin, Farahdiba and Mutter, Didier and Padoy, Nicolas},
  journal={International Journal of Computer Assisted Radiology and Surgery},
  volume={21},
  number={1},
  pages={93--102},
  year={2026},
  publisher={Springer}
}

@article{mofo,
  title={Multi-organ foundation model for universal ultrasound image segmentation with task prompt and anatomical prior},
  author={Chen, Haobo and Cai, Yehua and Wang, Changyan and Chen, Lin and Zhang, Bo and Han, Hong and Guo, Yuqing and Ding, Hong and Zhang, Qi},
  journal={IEEE Transactions on Medical Imaging},
  volume={44},
  number={2},
  pages={1005--1018},
  year={2024},
  publisher={IEEE}
}

@article{perceptguide,
  title={An orchestration learning framework for ultrasound imaging: Prompt-guided hyper-perception and attention-matching downstream synchronization},
  author={Lin, Zehui and Li, Shuo and Wang, Shanshan and Gao, Zhifan and Sun, Yue and Lam, Chan-Tong and Hu, Xindi and Yang, Xin and Ni, Dong and Tan, Tao},
  journal={Medical Image Analysis},
  volume={104},
  pages={103639},
  year={2025},
  publisher={Elsevier}
}

@inproceedings{universeg,
  title={Universeg: Universal medical image segmentation},
  author={Butoi, Victor Ion and Ortiz, Jose Javier Gonzalez and Ma, Tianyu and Sabuncu, Mert R and Guttag, John and Dalca, Adrian V},
  booktitle={2023 IEEE/CVF international conference on computer vision (ICCV)},
  pages={21381--21394},
  year={2023},
  organization={IEEE}
}

@inproceedings{insid3,
  title     = {{INSID3}: Training-Free In-Context Segmentation with {DINOv3}},
  author    = {Claudia Cuttano and Gabriele Trivigno and Christoph Reich and Daniel Cremers and Carlo Masone and Stefan Roth},
  booktitle = {Proceedings of the IEEE/CVF Conference on Computer Vision and Pattern Recognition (CVPR)},
  year      = {2026}
}

@inproceedings{tyche,
  title={Tyche: Stochastic in-context learning for medical image segmentation},
  author={Rakic, Marianne and Wong, Hallee E and Ortiz, Jose Javier Gonzalez and Cimini, Beth A and Guttag, John V and Dalca, Adrian V},
  booktitle={2024 IEEE/CVF Conference on Computer Vision and Pattern Recognition (CVPR)},
  pages={11159--11173},
  year={2024},
  organization={IEEE}
}

@article{wang2022boundary,
  title={Boundary-aware context neural network for medical image segmentation},
  author={Wang, Ruxin and Chen, Shuyuan and Ji, Chaojie and Fan, Jianping and Li, Ye},
  journal={Medical image analysis},
  volume={78},
  pages={102395},
  year={2022},
  publisher={Elsevier}
}

@article{A1,
  title={Foundation models for generalist medical artificial intelligence},
  author={Moor, Michael and Banerjee, Oishi and Abad, Zahra Shakeri Hossein and Krumholz, Harlan M and Leskovec, Jure and Topol, Eric J and Rajpurkar, Pranav},
  journal={Nature},
  volume={616},
  number={7956},
  pages={259--265},
  year={2023},
  publisher={Nature Publishing Group UK London}
}

@article{A2,
  title={Large ai models in health informatics: Applications, challenges, and the future},
  author={Qiu, Jianing and Li, Lin and Sun, Jiankai and Peng, Jiachuan and Shi, Peilun and Zhang, Ruiyang and Dong, Yinzhao and Lam, Kyle and Lo, Frank P-W and Xiao, Bo and others},
  journal={IEEE Journal of Biomedical and Health Informatics},
  volume={27},
  number={12},
  pages={6074--6087},
  year={2023},
  publisher={IEEE}
}

@article{A3,
  title={Mis-fm: 3d medical image segmentation using foundation models pretrained on a large-scale unannotated dataset},
  author={Wang, Guotai and Wu, Jianghao and Luo, Xiangde and Liu, Xinglong and Li, Kang and Zhang, Shaoting},
  journal={arXiv preprint arXiv:2306.16925},
  year={2023}
}

@article{A4,
  title={A foundation model for generalizable disease detection from retinal images},
  author={Zhou, Yukun and Chia, Mark A and Wagner, Siegfried K and Ayhan, Murat S and Williamson, Dominic J and Struyven, Robbert R and Liu, Timing and Xu, Moucheng and Lozano, Mateo G and Woodward-Court, Peter and others},
  journal={Nature},
  volume={622},
  number={7981},
  pages={156--163},
  year={2023},
  publisher={Nature Publishing Group UK London}
}

@inproceedings{A5,
  title={Foundation model for endoscopy video analysis via large-scale self-supervised pre-train},
  author={Wang, Zhao and Liu, Chang and Zhang, Shaoting and Dou, Qi},
  booktitle={International conference on medical image computing and computer-assisted intervention},
  pages={101--111},
  year={2023},
  organization={Springer}
}

@article{A6,
  title={URFM: A general Ultrasound Representation Foundation Model for advancing ultrasound image diagnosis},
  author={Kang, Qingbo and Lao, Qicheng and Gao, Jun and Bao, Wuyongga and He, Zhu and Du, Chenlin and Lu, Qiang and Li, Kang},
  journal={IScience},
  volume={28},
  number={8},
  year={2025},
  publisher={Elsevier}
}

@article{A7,
  title={Tinyusfm: Towards compact and efficient ultrasound foundation models},
  author={Ma, Chen and Jiao, Jing and Liang, Shuyu and Fu, Junhu and Wang, Qin and Li, Zeju and Wang, Yuanyuan and Guo, Yi},
  journal={IEEE Journal of Biomedical and Health Informatics},
  year={2026},
  publisher={IEEE}
}

@inproceedings{C1,
  title={Seggpt: Towards segmenting everything in context},
  author={Wang, Xinlong and Zhang, Xiaosong and Cao, Yue and Wang, Wen and Shen, Chunhua and Huang, Tiejun},
  booktitle={2023 IEEE/CVF International Conference on Computer Vision (ICCV)},
  pages={1130--1140},
  year={2023},
  organization={IEEE}
}

@inproceedings{C2,
  title={Personalize segment anything model with one shot},
  author={Zhang, Renrui and Jiang, Zhengkai and Guo, Ziyu and Yan, Shilin and Pan, Junting and Dong, Hao and Qiao, Yu and Peng, Gao and Li, Hongsheng},
  booktitle={International Conference on Learning Representations},
  volume={2024},
  pages={18250--18279},
  year={2024}
}

@article{vit,
  title={An image is worth 16x16 words: Transformers for image recognition at scale},
  author={Dosovitskiy, Alexey},
  journal={arXiv preprint arXiv:2010.11929},
  year={2020}
}

@article{breast,
  title={Curated benchmark dataset for ultrasound based breast lesion analysis},
  author={Paw{\l}owska, Anna and {\'C}wierz-Pie{\'n}kowska, Anna and Domalik, Agnieszka and Jagu{\'s}, Dominika and Kasprzak, Piotr and Matkowski, Rafa{\l} and Fura, {\L}ukasz and Nowicki, Andrzej and {\.Z}o{\l}ek, Norbert},
  journal={Scientific Data},
  volume={11},
  number={1},
  pages={148},
  year={2024},
  publisher={Nature Publishing Group UK London}
}

@article{busbra,
  title={BUS-BRA: a breast ultrasound dataset for assessing computer-aided diagnosis systems},
  author={G{\'o}mez-Flores, Wilfrido and Gregorio-Calas, Maria Julia and Coelho de Albuquerque Pereira, Wagner},
  journal={Medical physics},
  volume={51},
  number={4},
  pages={3110--3123},
  year={2024},
  publisher={Wiley Online Library}
}

@article{busi,
  title={Dataset of breast ultrasound images},
  author={Al-Dhabyani, Walid and Gomaa, Mohammed and Khaled, Hussien and Fahmy, Aly},
  journal={Data in brief},
  volume={28},
  pages={104863},
  year={2020},
  publisher={Elsevier}
}

@article{busi-whu,
  title={DFLNet: Disentangled feature learning network for breast cancer ultrasound image segmentation},
  author={Ye, Zhaoyi and Huang, Jin and Zhang, Yimin and Deng, Jingwen and Zhang, Jingwen and Liu, Sheng and Wang, Du and Mei, Liye and Lei, Cheng},
  journal={Digital Signal Processing},
  volume={165},
  pages={105331},
  year={2025},
  publisher={Elsevier}
}

@inproceedings{ddti,
  title={An open access thyroid ultrasound image database},
  author={Pedraza, Lina and Vargas, Carlos and Narv{\'a}ez, Fabi{\'a}n and Dur{\'a}n, Oscar and Mu{\~n}oz, Emma and Romero, Eduardo},
  booktitle={10th International symposium on medical information processing and analysis},
  volume={9287},
  pages={188--193},
  year={2015},
  organization={SPIE}
}

@article{tn3ktg3k,
  title={Thyroid region prior guided attention for ultrasound segmentation of thyroid nodules},
  author={Gong, Haifan and Chen, Jiaxin and Chen, Guanqi and Li, Haofeng and Li, Guanbin and Chen, Fei},
  journal={Computers in biology and medicine},
  volume={155},
  pages={106389},
  year={2023},
  publisher={Elsevier}
}

@article{camus,
  title={Deep learning for segmentation using an open large-scale dataset in 2D echocardiography},
  author={Leclerc, Sarah and Smistad, Erik and Pedrosa, Joao and {\O}stvik, Andreas and Cervenansky, Frederic and Espinosa, Florian and Espeland, Torvald and Berg, Erik Andreas Rye and Jodoin, Pierre-Marc and Grenier, Thomas and others},
  journal={IEEE transactions on medical imaging},
  volume={38},
  number={9},
  pages={2198--2210},
  year={2019},
  publisher={IEEE}
}

@article{luminous,
  title={LUMINOUS database: lumbar multifidus muscle segmentation from ultrasound images},
  author={Belasso, Clyde J and Behboodi, Bahareh and Benali, Habib and Boily, Mathieu and Rivaz, Hassan and Fortin, Maryse},
  journal={BMC musculoskeletal disorders},
  volume={21},
  number={1},
  pages={703},
  year={2020},
  publisher={Springer}
}

@article{psfhs,
  title={PSFHS: intrapartum ultrasound image dataset for AI-based segmentation of pubic symphysis and fetal head},
  author={Chen, Gaowen and Bai, Jieyun and Ou, Zhanhong and Lu, Yaosheng and Wang, Huijin},
  journal={Scientific data},
  volume={11},
  number={1},
  pages={436},
  year={2024},
  publisher={Nature Publishing Group UK London}
}

@article{usd514,
  title={Query2: Query over queries for improving gastrointestinal stromal tumour detection in an endoscopic ultrasound},
  author={He, Qi and Bano, Sophia and Liu, Jing and Liu, Wentian and Stoyanov, Danail and Zuo, Siyang},
  journal={Computers in Biology and Medicine},
  volume={152},
  pages={106424},
  year={2023},
  publisher={Elsevier}
}

@article{bus-uclm,
  title={BUS-UCLM: Breast ultrasound lesion segmentation dataset},
  author={Vallez, Noelia and Bueno, Gloria and Deniz, Oscar and Rienda, Miguel Angel and Pastor, Carlos},
  journal={Scientific Data},
  volume={12},
  number={1},
  pages={242},
  year={2025},
  publisher={Nature Publishing Group UK London}
}

@article{udiat,
  title={Automated breast ultrasound lesions detection using convolutional neural networks},
  author={Yap, Moi Hoon and Pons, Gerard and Marti, Joan and Ganau, Sergi and Sentis, Melcior and Zwiggelaar, Reyer and Davison, Adrian K and Marti, Robert},
  journal={IEEE journal of biomedical and health informatics},
  volume={22},
  number={4},
  pages={1218--1226},
  year={2017},
  publisher={IEEE}
}

@article{fugc,
  title={Fugc: Benchmarking semi-supervised learning methods for cervical segmentation},
  author={Bai, Jieyun and Tang, Yitong and Zhou, Zihao and Islam, Mahdi and Tabassum, Musarrat and Almar-Munoz, Enrique and Liu, Hongyu and Meng, Hui and Lv, Nianjiang and Deng, Bo and others},
  journal={IEEE Transactions on Medical Imaging},
  year={2026},
  publisher={IEEE}
}

@article{microseg,
  title={MicroSegNet: A deep learning approach for prostate segmentation on micro-ultrasound images},
  author={Jiang, Hongxu and Imran, Muhammad and Muralidharan, Preethika and Patel, Anjali and Pensa, Jake and Liang, Muxuan and Benidir, Tarik and Grajo, Joseph R and Joseph, Jason P and Terry, Russell and others},
  journal={Computerized Medical Imaging and Graphics},
  volume={112},
  pages={102326},
  year={2024},
  publisher={Elsevier}
}

@article{mmotu,
  title={MMOTU: A multi-modality ovarian tumor ultrasound image dataset for unsupervised cross-domain semantic segmentation},
  author={Zhao, Qi and Lyu, Shuchang and Bai, Wenpei and Cai, Linghan and Liu, Binghao and Cheng, Guangliang and Wu, Meijing and Sang, Xiubo and Yang, Min and Chen, Lijiang},
  journal={arXiv preprint arXiv:2207.06799},
  year={2022}
}

@misc{neck,
    author = {Anna Montoya and Hasnin and kaggle446 and shirzad and Will Cukierski and yffud},
    title = {Ultrasound Nerve Segmentation},
    year = {2016},
    howpublished = {\url{https://kaggle.com/competitions/ultrasound-nerve-segmentation}},
    note = {Kaggle}
}

@inproceedings{kidneyus,
  title={The open kidney ultrasound data set},
  author={Singla, Rohit and Ringstrom, Cailin and Hu, Grace and Lessoway, Victoria and Reid, Janice and Nguan, Christopher and Rohling, Robert},
  booktitle={International Workshop on Advances in Simplifying Medical Ultrasound},
  pages={155--164},
  year={2023},
  organization={Springer}
}

@misc{tnscui,
  title        = {Thyroid Nodule Segmentation and Classification in Ultrasound Images},
  author       = {Zhou, Jianqiao and Jia, Xiaohong and Ni, Dong and Noble, Alison and Huang, Ruobing and Tan, Tao and Van, Manh The},
  year         = 2020,
  month        = mar,
  publisher    = {Zenodo},
  doi          = {10.5281/zenodo.3715942},
  url          = {https://doi.org/10.5281/zenodo.3715942}
}

@article{nnunet,
  title={nnU-Net: a self-configuring method for deep learning-based biomedical image segmentation},
  author={Isensee, Fabian and Jaeger, Paul F and Kohl, Simon AA and Petersen, Jens and Maier-Hein, Klaus H},
  journal={Nature methods},
  volume={18},
  number={2},
  pages={203--211},
  year={2021},
  publisher={Nature Publishing Group US New York}
}

@article{tip0,
  title={Style consistency unsupervised domain adaptation medical image segmentation},
  author={Chen, Lang and Bian, Yun and Zeng, Jianbin and Meng, Qingquan and Zhu, Weifang and Shi, Fei and Shao, Chengwei and Chen, Xinjian and Xiang, Dehui},
  journal={IEEE Transactions on Image Processing},
  volume={33},
  pages={4882--4895},
  year={2024},
  publisher={IEEE}
}

@article{tip1,
  title={Cross-domain few-shot medical image segmentation via dynamic semantic matching},
  author={Zhu, Yazhou and Wang, Shidong and Zhou, Tao and Li, Zechao and Zhang, Haofeng and Shao, Ling},
  journal={IEEE Transactions on Image Processing},
  year={2025},
  publisher={IEEE}
}

@article{tip2,
  title={Learning shape-invariant representation for generalizable semantic segmentation},
  author={Zhang, Yuhang and Tian, Shishun and Liao, Muxin and Hua, Guoguang and Zou, Wenbin and Xu, Chen},
  journal={IEEE Transactions on Image Processing},
  volume={32},
  pages={5031--5045},
  year={2023},
  publisher={IEEE}
}

@article{tip3,
  title={Mdf-net: A multi-scale dynamic fusion network for breast tumor segmentation of ultrasound images},
  author={Qi, Wenbo and Wu, Ho-Chun and Chan, Shing-Chow},
  journal={IEEE Transactions on Image Processing},
  volume={32},
  pages={4842--4855},
  year={2023},
  publisher={IEEE}
}

\end{document}